%% file: main.tex
\documentclass[10pt]{article} 
\usepackage[preprint]{tmlr}

\input{math_commands.tex}

\usepackage[utf8]{inputenc} 
\usepackage{url}            
\usepackage{booktabs}       
\usepackage{amsfonts}       
\usepackage{nicefrac}       
\usepackage{microtype}      
\usepackage{diagbox}        
\usepackage{makecell}
\usepackage{xspace}
\usepackage{url}
\usepackage{graphicx}
\usepackage{bm}
\usepackage{amssymb}
\usepackage{amsmath}
\usepackage{amsthm}
\usepackage{algorithm}
\usepackage{algpseudocode}
\usepackage[table,xcdraw,dvipsnames]{xcolor}
\usepackage[title]{appendix}
\usepackage{wrapfig}
\usepackage{hyperref}
\definecolor{customblue}{rgb}{0.36, 0.55, 0.75}
\definecolor{myred}{RGB}{220,0,0}
\definecolor{myblue}{RGB}{0,80,255}
\definecolor{mygreen}{RGB}{0,150,0}
\usepackage{comment}
\usepackage{bbm}
\usepackage{multirow}
\usepackage{float}
\usepackage{subcaption}
\usepackage{makecell}

\newcommand{\cut}[1]{}
\newcommand{\std}[1]{\scriptsize{$\pm$#1}}

\newcommand{\myname}{ERP-FM\xspace}

\usepackage{listings}
\usepackage{graphicx}
\usepackage{amsmath}
\usepackage{setspace}
\usepackage{pifont}
\usepackage{tikz}

\title{
    ERP-FM: A Foundation Model for Universal ERP Analysis
}

\author{\name Yihe Wang \email ywang145@charlotte.edu \\
      University of North Carolina at Charlotte
      \AND
      \name Bohan Chen \email bchen130@jh.edu \\
      Johns Hopkins University
      \AND
      \name Taida Li \email tli14@charlotte.edu\\
      University of North Carolina at Charlotte
      \AND
      \name Yujun Yan \email yujun.yan@dartmouth.edu\\
      Dartmouth College
      \AND
      \name Rui Yin \email ruiyin@ufl.edu\\
      University of Florida
      \AND
      \name Xiang Zhang \email xiang.zhang@charlotte.edu\\
      University of North Carolina at Charlotte
}

\def\month{MM}  
\def\year{YYYY} 
\def\openreview{\url{https://openreview.net/forum?id=XXXX}} 

\begin{document}

\maketitle

\begin{abstract}

\input{000abstract.tex}

\end{abstract}

\section{Introduction}
\label{sec:intro}

\input{010intro}

\section{Related Work}
\label{sec:related}

\input{020related}

\section{Datasets}
\label{sec:preliminary}

\input{030preliminaries}

\section{Method}
\label{sec:method}

\input{040method}

\section{Experiments}
\label{sec:experiments}

\input{050setup}

\input{060results}

\section{Conclusion and Limitations}
\label{sec:conclusion}

\input{070conclusion}

\clearpage
\newpage

\bibliography{refs}
\bibliographystyle{tmlr}


\clearpage

\appendix
\setcounter{page}{1}
\begin{appendices}

\input{080appendix}

\end{appendices}


\clearpage

\end{document}

%% file: math_commands.tex
\usepackage{amsmath,amsfonts,bm}

\def\eqref#1{equation~\ref{#1}}

\def\1{\bm{1}}

\DeclareMathAlphabet{\mathsfit}{\encodingdefault}{\sfdefault}{m}{sl}
\SetMathAlphabet{\mathsfit}{bold}{\encodingdefault}{\sfdefault}{bx}{n}



%% file: 000abstract.tex
Foundation models have recently shown strong potential for learning generalizable EEG representations, yet their effectiveness for event-related potential (ERP) analysis remains unclear. In this work, we investigate two fundamental questions: 1) can foundation-model learning benefit ERP analysis, and what limits the transfer of existing EEG foundation models to ERP tasks? 2) can the complementary advantages of single-trial and averaged-trial ERP be integrated into a unified training pipeline? To study these questions, we curate a large-scale ERP corpus comprising 1,517,157 single-trial ERPs from 3,696 subjects across 38 datasets and 18 paradigms. Leveraging this corpus, we introduce \myname, to the best of our knowledge, the first foundation model specifically developed for ERP representation learning. \myname uses single-channel tokenization, temporal and spatial positional embeddings, and mixed masked autoencoding for large-scale single-trial pretraining. We compare our model against 17 existing methods on 12 downstream datasets covering ERP event/condition classification and neurological disease classification. Our model achieves the best overall average rank across all evaluated methods. Furthermore, our analyses reveal that both ERP and non-ERP EEG pretraining can provide transferable representations for ERP tasks, while fine-grained temporal tokenization is critical for effectively modeling transient ERP dynamics. We further find that single-trial and averaged-trial ERP play complementary rather than competing roles. Combining single-trial pretraining with averaged-trial downstream adaptation substantially improves neurological disease analysis. Overall, these findings establish an effective foundation-model training pipeline for ERP analysis and represent significant progress toward generalizable ERP representation learning. Source code: \url{https://github.com/DL4mHealth/ERP-FM}

%% file: 010intro.tex
ERP is a transient electrophysiological response embedded in ongoing electroencephalography (EEG) and associated with discrete sensory, cognitive, or behavioral events~\citep{sur2009event}. Different ERP components reflect a broad range of perceptual, cognitive, affective, and motor processes, making ERP analysis widely used in cognitive neuroscience~\citep{jackson2023reduced}, brain-computer interface (BCI)~\citep{chailloux2020single}, and neurological and psychiatric disease analysis~\citep{breitling2020economical,singh2023evoked}. Despite its broad applications, much of conventional ERP analysis remains highly hypothesis-driven and often relies on manually designed analysis procedures~\citep{kappenman2021erp,toffolo2022evoking}. Researchers typically predefine ERP components, electrodes, and latency windows, average trials within experimental conditions, and quantify handcrafted characteristics such as component amplitude and latency before statistical analysis or classification~\citep{rehman2025machine,batbat2026multimodal}. Although these approaches provide strong neurophysiological interpretability, they require substantial domain expertise and task-specific design, making it difficult to develop a unified analysis framework across heterogeneous ERP paradigms. These limitations motivate more general data-driven approaches that can learn ERP representations directly from signals with reduced reliance on manually specified features.

Deep learning provides an alternative paradigm by learning representations directly from EEG signals, substantially reducing the dependence on manually designed features~\citep{lawhern2018eegnet,song2022eeg,ding2024eeg}. More recently, EEG foundation models have extended this idea through large-scale self-supervised pretraining, aiming to learn generalizable neural representations that can transfer across datasets, subjects, and downstream tasks. Recent models such as NeurIPT~\citep{fang2025neuript}, DeeperBrain~\citep{wang2026deeperbrain}, and ST-EEGFormer~\citep{yang2026eeg} demonstrate the rapidly growing capability of large-scale EEG representation learning. However, existing foundation models are predominantly developed and pretrained on continuous or non-ERP EEG recordings, \footnote{EEG paradigms do not follow a strict binary taxonomy based on time or phase locking. Resting-state EEG has no external event reference; cue-based motor imagery is event-related but is commonly characterized by induced, non-phase-locked ERD/ERS; periodic evoked responses such as SSVEP can be strongly phase-locked despite not being transient ERPs. For simplicity, we use \textbf{non-ERP EEG} throughout this work to denote EEG recordings that are not analyzed as transient ERP responses.} while ERP-specific characteristics have rarely been considered as a primary design objective. Consistent with this gap, ERP-Benchmark~\citep{wang2026benchmarking} showed that existing EEG foundation models can even underperform standard deep learning models on ERP tasks, in contrast to the strong performance gains observed from foundation models in other domains such as natural language processing (NLP)~\citep{liu2024deepseek} and computer vision (CV)~\citep{simeoni2025dinov3}. This motivates the first question of this work: \textit{can foundation-model learning benefit ERP analysis, and if so, why do existing EEG foundation models transfer poorly to ERP tasks?}

Beyond representation learning itself, ERP analysis also presents a second fundamental challenge: the trade-off between \textbf{single-trial} and \textbf{averaged-trial} analysis. Conventional ERP studies commonly average repeated trials from the same subject and experimental event/condition to suppress non-phase-locked background activity and improve the SNR of event-related responses~\citep{mouraux2008across}. Averaged-trial ERPs provide cleaner and more interpretable waveforms, making differences in component morphology, amplitude, and latency easier to characterize~\citep{lazzari2024beyond,quattrociocchi2026improving}. However, trial averaging discards trial-to-trial variability and can drastically reduce the number of available training samples. In contrast, single-trial ERP analysis preserves individual responses and provides substantially more samples for data-driven learning and real-time BCI applications~\citep{chailloux2020single,geng2026zmw}, but each trial contains substantially more background EEG activity and noise. This creates a natural complementarity: single-trial ERP provides the scale needed for representation learning, whereas averaged-trial ERP provides higher signal quality and neurophysiological interpretability. This motivates the second question of this work: \textit{can a foundation-model framework combine the advantages of single-trial and averaged-trial ERP analysis for practical downstream applications?}

\begin{figure*}[t]
    \centering    
    \includegraphics[width=1.0\linewidth]{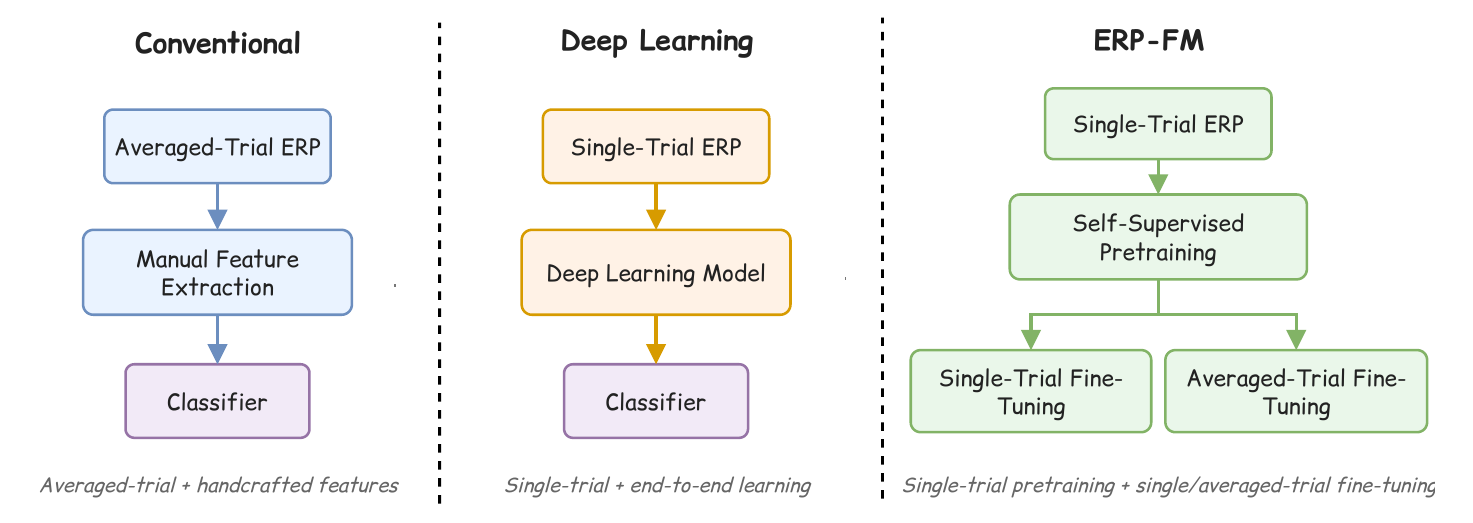}
    \caption{\textbf{Framework Comparison.} Our pipeline and commonly used ERP analysis pipelines.
    }
    \label{fig:framework_comparison}
    \vspace{-5mm}
\end{figure*}

To answer these questions, we introduce \textbf{\myname}, to the best of our knowledge, the first foundation model specifically developed and pretrained at scale for ERP representation learning. We construct a large-scale pretraining corpus containing \textbf{1,517,157 single-trial ERPs from 3,696 subjects across 38 datasets and 18 ERP paradigms} and pretrain a lightweight Transformer using masked autoencoding~\citep{he2022masked}. We systematically evaluate \myname on 12 downstream datasets covering ERP cognitive event/condition classification and ERP-based neurological disease detection. \myname achieves the best overall ranking among 18 evaluated methods, with Top-1 performance in 20 of 36 dataset-metric evaluations, demonstrating that large-scale self-supervised pretraining can effectively support ERP decoding. Beyond establishing its performance, we use \myname as a controlled framework to investigate why existing EEG foundation models may transfer poorly to ERP. We find that non-ERP EEG pretraining also provides transferable representations, suggesting that the limitation cannot be explained by pretraining data modality alone. In contrast, temporal resolution has a substantial effect: 0.25- and 0.5-second embedding patches consistently outperform a 1-second patch, suggesting that the approximately 1-second temporal units adopted by models such as BIOT~\citep{yang2024biot}, LaBraM~\citep{jiang2024large}, CBraMod~\citep{wang2024cbramod}, and REVE~\citep{el2026reve} may be too coarse for fine-grained ERP dynamics. Finally, self-supervised pretraining and conventional trial averaging complement each other strongly. We demonstrate a pipeline in which single-trial self-supervised pretraining followed by averaged-trial downstream fine-tuning and subject-level majority voting provides the strongest overall performance for ERP-based neurological disease detection, while retaining averaged ERP waveforms that are naturally suited for subsequent future neurophysiological interpretation. Figure~\ref{fig:framework_comparison} illustrates our pipeline and the commonly used pipelines in conventional feature extraction and deep learning methods for ERP data analysis.

Our main contributions are summarized as follows: 1) Large-scale ERP pretraining corpus. We curate a large-scale ERP pretraining resource from open-access datasets, forming the largest and most diverse ERP pretraining corpus to date, to the best of our knowledge. 2) First ERP foundation model. We introduce \textbf{\myname}, a lightweight ERP foundation model that achieves SOTA performance and outperforms existing EEG foundation models and strong supervised deep learning methods. 3) Understanding ERP foundation-model design. We show that non-ERP EEG contains transferable information for ERP tasks, while coarse temporal tokenization may be a key factor limiting ERP decoding performance. 4) Bridging single-trial and averaged-trial ERP analysis. We demonstrate that the complementary advantages of single-trial and averaged-trial ERP can be effectively combined into a unified pipeline for ERP-based downstream applications, particularly neurological disease detection.

%% file: 020related.tex
\textbf{Event-Related Potential Analysis.}
\label{para:erp_analysis}
ERP analysis is widely used to study cognitive mechanisms, BCI decoding, and neurological and psychiatric disorders. ERP CORE~\citep{kappenman2021erp} provides standardized paradigms and analysis pipelines for seven widely studied ERP components, supporting reproducible investigations of sensory, cognitive, and motor processes. CESCA~\citep{isbell2025cognitive} collects multiple cognitive ERP paradigms to investigate cognitive electrophysiology in adulthood across different socioeconomic contexts, while the Nencki-Symfonia ERP dataset~\citep{dzianok2022nencki} combines multiple cognitive tasks and resting-state recordings for broad cognitive electrophysiology research. ERP responses to semantically congruent and incongruent sentences are used to characterize semantic processing~\citep{toffolo2022evoking}, while reward-related ERP studies investigate reinforcement learning and affective processing through reward positivity and related electrophysiological responses~\citep{jackson2023reduced,brown2022reward}. ERP also provides useful biomarkers for neurological and psychiatric analysis. Working-memory and response-inhibition ERPs characterize cognitive deficits in ADHD~\citep{breitling2020economical}, evoked mid-frontal activity reflects cognitive dysfunction in Parkinson's disease~\citep{singh2023evoked}, and ERP responses predict symptomatic distress and recovery following mild traumatic brain injury~\citep{cavanagh2019erps}. Recent machine-learning approaches continue to extract predefined ERP components, including amplitude and latency features, for downstream classification~\citep{rehman2025machine,batbat2026multimodal}, while single-trial deep learning directly decodes event-related responses for real-time BCI applications~\citep{geng2026zmw}. ERP-Benchmark provides a comprehensive evaluation of handcrafted features, supervised deep learning methods, and foundation models for single-trial ERP classification~\citep{wang2026benchmarking}. Overall, existing ERP studies cover diverse scientific and clinical applications, but their analysis pipelines are generally developed for individual paradigms, ERP components, or downstream tasks, motivating more generalizable representation learning across heterogeneous ERP datasets.

\textbf{EEG Foundation Models.}
\label{para:eeg_foundation_model}
EEG foundation models increasingly leverage large-scale self-supervised pretraining to learn transferable representations across heterogeneous EEG datasets and downstream tasks. BIOT~\citep{yang2024biot} develops a Transformer-based biosignal representation framework for cross-data learning across heterogeneous recording configurations using single-channel patch embedding. LaBraM~\citep{jiang2024large} scales self-supervised EEG pretraining to large collections of EEG recordings and learns generic representations for diverse BCI tasks. CBraMod~\citep{wang2024cbramod} introduces a criss-cross Transformer to explicitly model spatial and temporal dependencies in multichannel EEG. NeurIPT~\citep{fang2025neuript} incorporates a progressive mixture-of-experts architecture and 3D electrode-coordinate embeddings to improve representation learning across heterogeneous neural interfaces. REVE~\citep{el2026reve} further scales EEG foundation-model pretraining to approximately 60,000 hours of recordings from 25,000 subjects and develops flexible representations that adapt to different recording configurations. Application-oriented foundation models also emerge for specific clinical domains; LEAD~\citep{wang2025lead} focuses large-scale EEG representation learning on Alzheimer's disease detection. DeeperBrain~\citep{wang2026deeperbrain} incorporates neurophysiological inductive biases into EEG representation learning, while ST-EEGFormer~\citep{yang2026eeg} demonstrates that a simple Transformer with raw-signal masked autoencoding can achieve strong performance across diverse EEG tasks. ECHO~\citep{liu2026echo} moves beyond conventional encoder-centric EEG foundation models by formulating EEG modeling as contextual sequence-to-sequence learning, enabling adaptation to heterogeneous tasks through in-context examples without parameter updates. Despite these advances, existing EEG foundation models primarily target general EEG and BCI applications, while large-scale representation learning specifically designed for ERP signals remains largely unexplored.

%% file: 030preliminaries.tex
\subsection{Datasets Selection}
\label{sub:dataset_selection}
Our downstream evaluation datasets largely follow those used in ERP-Benchmark, except that \textbf{mTBI-ODD} is moved from downstream evaluation to the pretraining corpus. Specifically, we include 6 datasets for ERP cognitive event/condition classification: \textbf{CESCA-AOOD}, \textbf{CESCA-VOOD}, and \textbf{CESCA-FLANKER}~\citep{isbell2025cognitive}; \textbf{TDBrain-ODD}\footnote{Starting from Version 3.1, TDBrain provides oddball ERP recordings for a subset of participants as a beta prerelease.}~\citep{van2022two}; and \textbf{NSERP-MSIT} and \textbf{NSERP-ODD}~\citep{dzianok2022nencki}. These datasets involve classifying cognitive or behavioral conditions associated with ERP-eliciting events, including different stimulus types, response types, and experimental conditions, such as target vs. standard stimuli in an oddball paradigm. We additionally include 6 datasets for ERP-based neurological disease classification: \textbf{ADHD-WMRI}~\citep{breitling2020economical}, \textbf{PD-SIM} and \textbf{PD-ODD}~\citep{singh2023evoked}, \textbf{SCPD}~\citep{singh2018mid}, \textbf{RLPD}~\citep{brown2020eeg}, and \textbf{AOPD}~\citep{cavanagh2021eeg}.  Examples include distinguishing subjects with Parkinson's disease (PD) from healthy controls (HC). All downstream datasets are guaranteed to have a sufficient number of subjects (more than 40) and clearly distinguishable labels, comprising a total of \textbf{382,121 trials from 434 unique subjects}.

To construct the pretraining corpus for the ERP foundation model, we systematically searched publicly accessible EEG repositories, including OpenNeuro, OSF, Figshare, and Dryad. We identified 38 ERP datasets covering \textbf{18 ERP paradigms}, comprising a total of \textbf{1,517,157 trials from 3,696 unique subjects}. To the best of our knowledge, this represents the largest and most diverse ERP corpus assembled for foundation-model pretraining to date. Notably, some datasets correspond to different ERP tasks collected from the same group of participants, such as \textbf{ERPCORE-ERN} and \textbf{ERPCORE-LRP}~\citep{kappenman2021erp}. We therefore carefully reviewed the associated publications and participant information to identify shared participant cohorts and ensure that \textbf{no subjects overlap between the pretraining corpus and the downstream evaluation datasets}. This strict subject separation prevents information leakage from pretraining into downstream evaluation. Moreover, to investigate whether non-ERP EEG data provide transferable representations for ERP downstream tasks, we additionally include 5 non-ERP EEG datasets in the pretraining ablation study, comprising a total of \textbf{2,544,890 trials from 3,709 unique subjects}, providing a subject cohort comparable in size to the ERP corpus and more training trials.

\input{tables/datasets/downstream_datasets}

\subsection{Data Preprocessing}
\label{sub:data_preprocess}
We follow the preprocessing pipeline established in ERP-Benchmark~\citep{wang2026benchmarking} to extract ERP/EEG epochs. \textbf{1) Removal of non-EEG channels:} All non-EEG channels, such as EOG channels and coordinate information, are removed. \textbf{2) Notch and band-pass filtering:} A notch filter at 50 Hz or 60 Hz is applied to suppress line noise, followed by a band-pass filter between 0.5 Hz and 45 Hz. \textbf{3) Bad channel interpolation:} Channels marked as bad are interpolated. \textbf{4) Average re-referencing:} Average re-referencing is applied to reduce dependence on the original recording reference. \textbf{5) Artifact removal:} Independent component analysis (ICA), combined with ICLabel~\cite{pion2019iclabel}, is used to automatically identify and remove components associated with eye blinks, muscle activity, and cardiac artifacts. \textbf{6) Resampling:} All recordings are resampled to a uniform sampling rate of 200 Hz. \textbf{7) Trial epoching:} For ERP datasets, EEG recordings are segmented into ERP trials according to stimulus, response, or feedback events, while recordings unrelated to ERP events, such as resting-state recordings, are discarded. For non-ERP EEG datasets, continuous recordings are segmented into fixed-length windows, such as 1- or 2-second segments. \textbf{8) Baseline correction:} Baseline correction is applied to ERP trials using the corresponding pre-event interval to reduce baseline offsets and emphasize event-related changes. This step is skipped for non-ERP EEG datasets. The baseline interval and epoch length vary across datasets. \textbf{9) Label assignment:} Each trial is assigned a neurological disease label $y$, a cognitive event/condition ID $e$, and a subject ID $s$. These labels are used for downstream classification, trial averaging, and subject-level disease detection via majority voting. Unavailable labels are assigned $-1$ as placeholders. \textbf{10) Metadata construction:} A metadata JSON file is created for each dataset to store information such as the channel montage, channel names, sampling rate, epoch boundaries, and label definitions for potential use in model design. \textbf{11) Z-score normalization:} During data loading, each channel of every segmented trial is independently normalized to zero mean and unit variance. \textbf{12) Trial averaging:} For averaged-trial analysis, all normalized trials sharing the same subject ID $s$ and ERP event/condition ID $e$ are averaged before downstream adaptation to obtain averaged-trial ERPs. This step applies only to experiments involving averaged-trial analysis. The statistics of the downstream datasets, ERP pretraining datasets, and non-ERP EEG pretraining datasets are summarized in Tables~\ref{tab:downstream_data},~\ref{tab:erp_pretraining_data}, and~\ref{tab:non_erp_pretraining_data}, respectively. We provide additional details on dataset-specific preprocessing in our GitHub repository.

%% file: tables/datasets/downstream_datasets.tex
\begin{table*}[t]
    \centering
    \caption{\textbf{Downstream Dataset Statistics.} All the downstream datasets follow the same preprocessing pipeline, including a band-pass filter, artifact removal, and resampling to 200Hz. The epoch and baseline window depend on the ERP tasks and datasets, following the reference paper of each dataset if provided. Abbreviations: \textbf{AODD}: Auditory Oddball; \textbf{VODD}: Visual Oddball; \textbf{MSIT+}: Extended multi-source interference task; \textbf{SIM}: Simon Conflict; \textbf{RL}: Reinforcement Learning; \textbf{HC}: Healthy Control; \textbf{PD}: Parkinson's Disease; \textbf{ADHD}: Attention Deficit Hyperactivity Disorder.
    }
    \vspace{-2mm}
    \label{tab:downstream_data}
    \resizebox{1.0\textwidth}{!}{%
    \begin{tabular}{@{}l|cccccccc@{}}
    \toprule
    \multicolumn{1}{l|}{\textbf{Datasets}}  & \textbf{ERP Task} & \textbf{\#Subjects} &  \textbf{Baseline(s)} &  \textbf{Epoch(s)}  & \textbf{\#Trials} & \textbf{\#Channels} & \textbf{Classes} & \textbf{Raw Data Link} \\ 
    \midrule
    \multicolumn{1}{l|}{\textbf{CESCA-AODD}} & AODD & 127 &  [-0.2, 0.0]  & [-0.2, 0.8] & 38,151 & 26 & Standard vs Target & \href{https://openneuro.org/datasets/ds006018/versions/1.2.2}{Openneuro} \\
    \multicolumn{1}{l|}{\textbf{CESCA-VODD}} & VODD & 127 &  [-0.2, 0.0]  & [-0.2, 0.8] & 20,419 & 26 & Standard vs Target & \href{https://openneuro.org/datasets/ds006018/versions/1.2.2}{Openneuro} \\
    \multicolumn{1}{l|}{\textbf{CESCA-FLANKER}} & Flanker & 73 &  [-0.2, 0.0]  & [-0.2, 0.8] & 29,774 & 26 &  Congruent vs Incongruent & \href{https://openneuro.org/datasets/ds006018/versions/1.2.2}{Openneuro} \\
    \multicolumn{1}{l|}{\textbf{TDBrain-ODD}} & Oddball & 127 &  [-0.2, 0.0]  & [-0.2, 0.8] & 54,400 & 26 & Standard vs Target & \href{https://brainclinics.com/resources}{Brainclinics} \\
    \multicolumn{1}{l|}{\textbf{NSERP-MSIT}} & MSIT+ & 42 &  [-0.5, 0.0]  & [-0.5, 1.0] & 16,729 & 123 & \makecell{Non-Conflict vs Simon Effect vs \\ Flanker Effect vs Double-Conflict} & \href{https://openneuro.org/datasets/ds004621/versions/1.0.4}{Openneuro} \\
    \multicolumn{1}{l|}{\textbf{NSERP-ODD}} & VODD & 42 &  [-0.5, 0.0]  & [-0.5, 1.0] & 27,865 & 123 & Standard vs Target vs Novel & \href{https://openneuro.org/datasets/ds004621/versions/1.0.4}{Openneuro} \\
    \multicolumn{1}{l|}{\textbf{PD-SIM}} & SIM & 147 &  [-0.3, -0.2]  & [-0.5, 1.0] & 55,921 & 60 & HC vs PD & \href{https://openneuro.org/datasets/ds004580/versions/1.0.0}{Openneuro} \\
    \multicolumn{1}{l|}{\textbf{PD-ODD}} & VODD & 146 & [-0.3, -0.2] & [-0.5, 1.0] & 34,702 & 60 & HC vs PD & \href{https://openneuro.org/datasets/ds004574/versions/1.0.0}{Openneuro} \\
    \multicolumn{1}{l|}{\textbf{ADHD-WMRI}} & N-Back, GoNogo  & 59 &  [-0.2, 0.0]  & [-0.2, 0.8] & 21,832 & 21 & HC vs ADHD & \href{https://figshare.com/articles/dataset/EEG_raw_data_-_Economical_Assessment_of_Working_Memory_and_Response_Inhibition_in_ADHD/12115773}{Figshare} \\
    \multicolumn{1}{l|}{\textbf{SCPD}} & SIM & 55 &  [-0.3, -0.2]  & [-0.5, 1.0] & 46,193 & 59 & HC vs PD & \href{https://openneuro.org/datasets/ds003509/versions/1.1.0}{Openneuro} \\
    \multicolumn{1}{l|}{\textbf{RLPD}} & RL & 56 &  [-0.2, 0.0]  & [-0.2, 0.8] & 21,510 & 59 & HC vs PD & \href{https://openneuro.org/datasets/ds003506/versions/1.1.0}{Openneuro} \\
    \multicolumn{1}{l|}{\textbf{AOPD}} & AODD & 50 &  [-0.2, 0.0]  & [-0.2, 0.8] & 14,625 & 59 & HC vs PD & \href{https://openneuro.org/datasets/ds003490/versions/1.1.0}{Openneuro} \\

    \bottomrule
    \end{tabular}
    } 
    \vspace{-5mm}
\end{table*}

%% file: 040method.tex
\begin{figure*}[t]
    \centering    \includegraphics[width=1.0\linewidth]{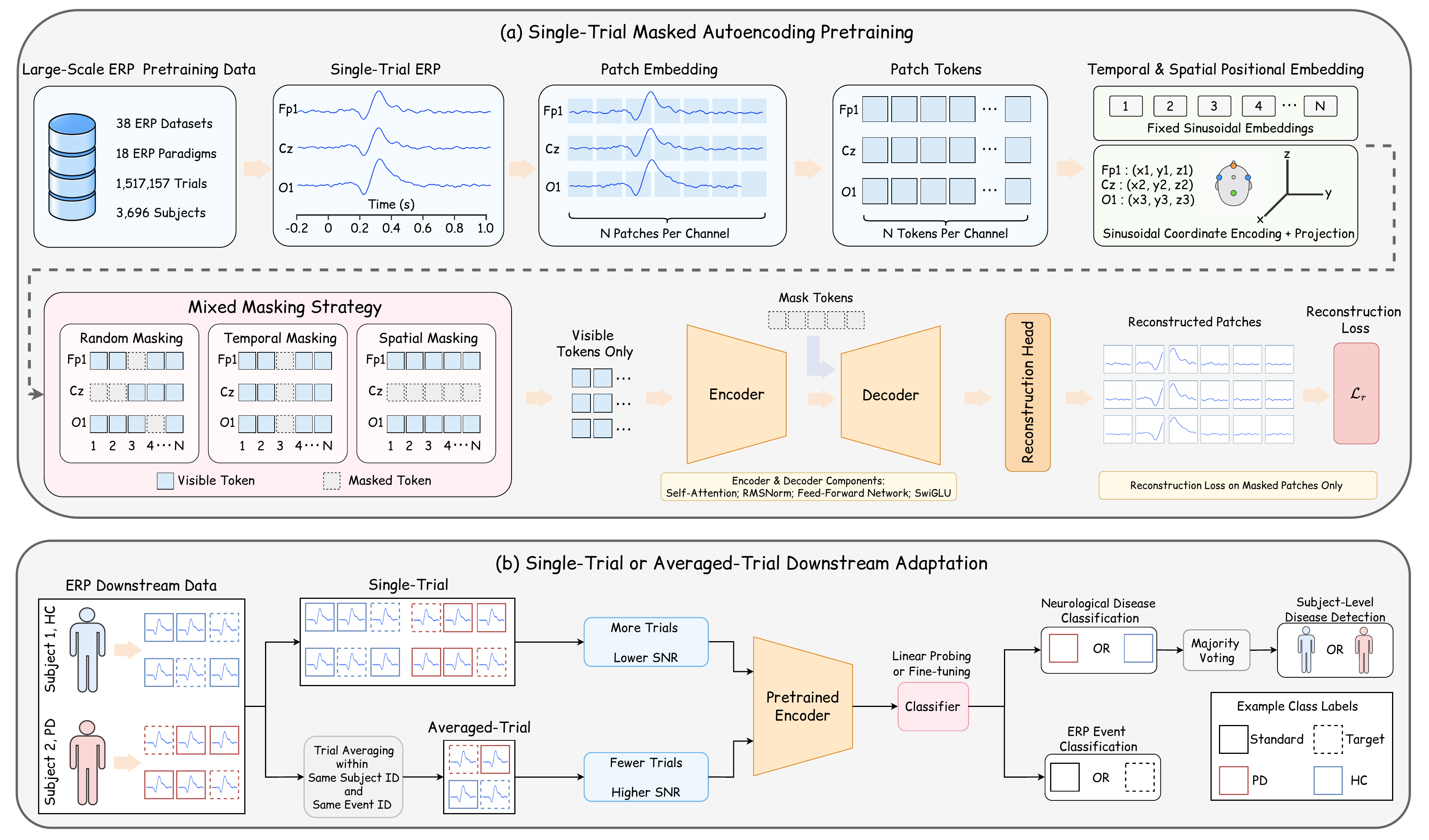}
    \caption{\textbf{ERP-FM Framework.} (a) \myname is pretrained on a large-scale single-trial ERP corpus using masked autoencoding. Each ERP epoch is divided into non-overlapping single-channel patches for tokenization and combined with temporal and spatial positional embeddings. A mixed masking strategy incorporating random, temporal, and spatial masking encourages the model to learn complementary temporal and spatial dependencies. We use an encoder-decoder architecture with a reconstruction head to reconstruct the masked signal patches. After pretraining, we discard the decoder and reconstruction head and retain only the pretrained encoder for downstream tasks. (b) For downstream adaptation, \myname supports both single-trial and averaged-trial ERP analysis through linear probing and fine-tuning. Averaged-trial ERPs are obtained by averaging trials from the same subject $s$ and ERP event/condition $e$, providing a higher SNR at the cost of fewer downstream samples. The framework supports ERP event/condition classification and neurological disease classification, with subject-level neurological disease detection further obtained through majority voting over predictions from the same subject.}
    \label{fig:erp_fm_framework}
    \vspace{-3mm}
\end{figure*}

\textbf{Framework Overview.}
\label{para:framework_overview}
Figure~\ref{fig:erp_fm_framework} illustrates the overall framework of \myname, which consists of self-supervised pretraining and downstream adaptation. During pretraining, \myname learns from a large-scale corpus of single-trial ERP data through masked autoencoding. Each ERP epoch is divided into non-overlapping single-channel patches and combined with temporal and spatial positional embeddings. A mixed masking strategy incorporating random, temporal, and spatial masking is applied, and an encoder-decoder Transformer reconstructs the masked signal patches. After pretraining, the decoder and reconstruction head are discarded, while the pretrained encoder is retained for downstream tasks. During downstream adaptation, the pretrained encoder can be utilized through linear probing or fine-tuning using either single-trial or averaged-trial ERP inputs. The framework supports ERP event/condition classification and neurological disease classification, with subject-level disease detection further obtained through majority voting.

\textbf{Patch Embedding.}
\label{para:patch_embedding}
To accommodate ERP trials with heterogeneous channel configurations and variable signal lengths, \myname adopts single-channel patch embedding. Given an ERP epoch $\bm{X}\in\mathbb{R}^{C\times T}$, where $C$ denotes the number of EEG channels and $T$ denotes the number of time samples, we independently divide each channel into non-overlapping univariate temporal patches of length $L$. Zero padding is applied along the temporal dimension when $T$ is not divisible by $L$, resulting in $N=\lceil T/L\rceil$ patches per channel. This yields a patchified tensor $\bm{X}_{p}\in\mathbb{R}^{C\times N\times L}$. Each patch is then projected into a $D$-dimensional token through a shared linear projector $\bm{W}\in\mathbb{R}^{L\times D}$:
\begin{equation}
    \bm{E} \leftarrow \bm{X}_{p}\bm{W},
    \quad \bm{E}\in\mathbb{R}^{C\times N\times D}.
\end{equation}
This single-channel patching strategy preserves local ERP waveform morphology within each electrode before the Transformer backbone models cross-channel interactions.

\textbf{Temporal and Spatial Positional Embeddings.}
\label{para:positional_embedding}
To make the patch tokens aware of both ERP latency and scalp location, \myname adds temporal and spatial positional embeddings to the patch embeddings. For temporal information, we use the fixed sinusoidal positional embeddings $\bm{TE}\in\mathbb{R}^{N\times D}$, following the original Transformer~\citep{vaswani2017attention}. For spatial information, we derive channel embeddings from electrode coordinates. During preprocessing, each dataset stores its channel montage and channel names in a metadata JSON file. Given the montage type, such as the international 10-20~\citep{homan1987cerebral} system, and channel names, such as Fp1 and Fp2, we retrieve the corresponding 3D electrode coordinates using MNE-Python~\citep{gramfort2013meg}. A sinusoidal coordinate encoding is then applied to the three spatial axes and concatenated into a $D$-dimensional spatial embedding, yielding $\bm{SE}\in\mathbb{R}^{C\times D}$~\citep{fang2025neuript}. The final input embeddings are obtained by broadcasting temporal embeddings over channels and spatial embeddings over temporal patches:
\begin{equation}
    \bm{E}
    \leftarrow
    \bm{E}
    +
    \operatorname{Broadcast}_{C}(\bm{TE})
    +
    \operatorname{Broadcast}_{N}(\bm{SE}),
    \quad
    \bm{E}\in\mathbb{R}^{C\times N\times D}.
\end{equation}
Here, $\operatorname{Broadcast}_{C}(\bm{TE})\in\mathbb{R}^{C\times N\times D}$ repeats temporal embeddings across channels, and $\operatorname{Broadcast}_{N}(\bm{SE})\in\mathbb{R}^{C\times N\times D}$ repeats spatial embeddings across temporal patches. In this way, each token is associated with both its temporal position within the ERP epoch and its physical electrode location on the scalp.

\textbf{Encoder-Decoder Transformer.}
\label{para:encoder_decoder_transformer}
An encoder-decoder Transformer processes the input patch embeddings $\bm{E}$. Following the recent simple EEG foundation model ST-EEGformer, we keep the backbone lightweight and standard. Each encoder/decoder layer consists of full self-attention~\citep{vaswani2017attention}, RMSNorm~\citep{zhang2019root}, and a feed-forward network with SwiGLU activation~\citep{shazeer2020glu}. During downstream adaptation, all tokens are fed into the encoder, while the decoder and reconstruction head are removed. A linear classifier is appended to the encoder and uses \textbf{all} output tokens for final classification. During pretraining, only visible tokens are fed into the encoder, while the decoder and reconstruction head reconstruct the original raw waveform patches at masked positions. The encoder is usually deeper than the decoder.

\textbf{Masking Strategies.}
\label{para:mask_strategy}
To improve the robustness of \myname to different types of missing information, we use three masking strategies during self-supervised pretraining: random, temporal, and spatial masking. Given the patchified tensor $\bm{X}_{p}\in\mathbb{R}^{C\times N\times L}$ and a mask ratio $r$, we define the masked patch set as $\Omega\subseteq\{1,\ldots,C\}\times\{1,\ldots,N\}$, where each pair $(c,n)$ corresponds to the $n$-th temporal patch of channel $c$. Random masking samples arbitrary patch positions from the full grid, following the standard MAE setting, with $|\Omega|\approx rCN$. Temporal masking samples a subset of temporal indices $\mathcal{N}_{\mathrm{mask}}\subset \{1,\ldots,N\}$ and masks them across all channels:
\begin{equation}
    \Omega
    =
    \{(c,n)\mid 1\leq c\leq C,\; n\in\mathcal{N}_{\mathrm{mask}}\},
    \quad |\mathcal{N}_{\mathrm{mask}}|\approx rN, \\
\end{equation}
which forces the model to infer potentially masked ERP components (e.g., P300, N400) from surrounding temporal context. Spatial masking samples a subset of channels $\mathcal{C}_{\mathrm{mask}}\subset \{1,\ldots,C\}$ and masks all temporal patches from these channels:
\begin{equation}
    \Omega
    =
    \{(c,n)\mid c\in\mathcal{C}_{\mathrm{mask}},\;1\leq n\leq N\},
    \quad |\mathcal{C}_{\mathrm{mask}}|\approx rC.
\end{equation}
which encourages the model to be robust to potentially missing electrodes and to exploit spatial relationships across channels. In practice, \myname can use either a fixed masking strategy or a mixed strategy, where one of the three strategies is randomly selected for each mini-batch.

\textbf{Single-Trial Masked Autoencoding Pretraining.}
\label{para:mae_pretraining}
Large-scale self-supervised pretraining benefits from a large number of training samples. Conventional ERP trial averaging improves signal quality by suppressing non-phase-locked background activity and noise, but it also collapses multiple repeated trials into a much smaller number of averaged ERPs. We therefore perform all self-supervised pretraining on \textbf{single-trial} ERP, preserving the substantially larger number of individual ERP observations available across datasets. Although individual ERP trials have a relatively low SNR, repeated single trials provide naturally variable and noisy realizations of the underlying event-related response. From a representation-learning perspective, such variability serves a role analogous to natural data augmentation, encouraging the model to learn ERP representations that are robust to background activity, noise, and trial-level variability.

\myname is pretrained with a raw-signal masked autoencoding objective~\citep{he2022masked}. For each single-trial ERP epoch, we first divide the signal into single-channel patches $\bm{X}_{p}\in\mathbb{R}^{C\times N\times L}$ and embed them into patch tokens $\bm{E}\in\mathbb{R}^{C\times N\times D}$. After applying one of the masking strategies described above, let $\Omega\subseteq\{1,\ldots,C\}\times\{1,\ldots,N\}$ denote the masked patch set and $\bar{\Omega}$ denote its visible complement. Only visible tokens $\{\bm{E}_{c,n}\mid(c,n)\in\bar{\Omega}\}$ are fed into the encoder. We denote the encoded visible representation at position $(c,n)$ as $\bm{H}_{c,n}$. The decoder restores the full channel-time token grid by filling masked positions with a shared learnable mask token $\bm{e}_{M}\in\mathbb{R}^{D}$. We first define the restored token grid $\bm{Z}^{0}\in\mathbb{R}^{C\times N\times D}$ as
\begin{equation}
    \bm{Z}^{0}_{c,n}
    =
    \begin{cases}
        \bm{H}_{c,n}, & (c,n)\in\bar{\Omega}, \\
        \bm{e}_{M}, & (c,n)\in\Omega.
    \end{cases}
\end{equation}
The decoder input is then obtained by adding the corresponding temporal and spatial positional embeddings to all restored tokens:
\begin{equation}
    \bm{Z}_{c,n}
    =
    \bm{Z}^{0}_{c,n}
    +
    \bm{TE}_{n}
    +
    \bm{SE}_{c},
    \quad
    1\leq c\leq C,\;1\leq n\leq N,
\end{equation}
where $\bm{TE}_{n},\bm{SE}_{c}\in\mathbb{R}^{D}$ denote the temporal and spatial positional embeddings for temporal patch $n$ and channel $c$, respectively. The decoder processes the restored token grid, and a reconstruction head maps the decoder outputs back to raw waveform patches:
\begin{equation}
    \hat{\bm{Y}}
    =
    \operatorname{Head}
    \left(
    \operatorname{Decoder}
    \left(
    \bm{Z}
    \right)
    \right),
    \quad
    \hat{\bm{Y}}\in\mathbb{R}^{C\times N\times L}.
\end{equation}

The reconstruction loss is computed only over masked patches:
\begin{equation}
    \mathcal{L}_{{r}}
    =
    \frac{1}{|\Omega|}
    \sum_{(c,n)\in\Omega}
    \ell
    \left(
    \hat{\bm{Y}}_{c,n},
    \bm{Y}_{c,n}
    \right),
\end{equation}
where $\bm{Y}=\bm{X}_{p}$ is the original raw patch target, $\bm{Y}_{c,n}\in\mathbb{R}^{L}$ is the original patch, $\hat{\bm{Y}}_{c,n}\in\mathbb{R}^{L}$ is the reconstructed patch, and $\ell(\cdot,\cdot)$ is the patch-level reconstruction loss. By default, we use smooth L1 loss, which has proven more robust to EEG pretraining in prior works~\citep{wang2026deeperbrain}. After pretraining, we discard the decoder and reconstruction head and retain only the pretrained encoder for downstream adaptation.

\textbf{Single-Trial or Averaged-Trial Downstream Adaptation.}
\label{para:downstream_adaptation}
Although single-trial ERP provides the large number of samples desirable for pretraining, downstream ERP analysis involves a different trade-off between sample quantity and signal quality. Single-trial ERP preserves the maximum number of downstream samples but contains substantial background activity and noise, whereas conventional trial averaging enhances phase-locked event-related responses and improves the SNR at the cost of substantially fewer samples. We therefore allow the pretrained \myname encoder to be adapted using either \textbf{single-trial} or \textbf{averaged-trial} ERP, enabling downstream tasks to benefit from either greater sample availability or cleaner ERP signals. To construct averaged-trial ERP, we group trials by subject ID $s$ and ERP event/condition ID $e$, and average all trials belonging to the same subject-event/condition pair. We then train on either single-trial or averaged-trial ERPs using the same encoder architecture. For linear probing, we freeze the pretrained encoder and optimize only the downstream classifier. For fine-tuning, we jointly optimize the pretrained encoder and classifier.

We consider two types of downstream tasks: ERP event/condition classification and neurological disease classification. For ERP event/condition classification, the classifier predicts the corresponding event or experimental condition $e$ of each ERP input. For neurological disease classification, the classifier predicts the disease label $y$ associated with each single-trial or averaged-trial ERP. Since neurological disease labels are defined at the subject level, we further obtain subject-level disease detection through majority voting~\citep{wang2025lead} over all predictions belonging to the same subject. In the single-trial setting, we vote over predictions from individual ERP trials, whereas in the averaged-trial setting, we vote over predictions from the event/condition-specific averaged ERPs for each subject.

%% file: 050setup.tex
\subsection{Experimental Setups}
\label{sub:experiment_setup}

\textbf{Baselines.}
We compare \myname with \textbf{17 baselines}, including 2 handcrafted-feature methods, 11 supervised deep learning models, and 4 EEG foundation models. These baselines are either SOTA methods or models that have demonstrated strong performance in prior EEG classification studies. The 2 handcrafted-feature methods use the same manual features as ERP-Benchmark~\cite{wang2026benchmarking}, followed by a linear classifier. The 11 supervised deep learning baselines include \textbf{TCN}~\citep{bai2018empirical}, \textbf{ModernTCN}~\citep{luo2024moderntcn}, \textbf{TimesNet}~\citep{wu2023timesnet}, \textbf{PatchTST}~\citep{nie2022time}, \textbf{iTransformer}~\citep{liuitransformer}, \textbf{Medformer}~\citep{wang2024medformer}, \textbf{MedGNN}~\citep{fan2025towards}, \textbf{EEGNet}~\citep{lawhern2018eegnet}, \textbf{EEGInception}~\citep{zhang2021eeg}, \textbf{EEGConformer}~\citep{song2022eeg}, and \textbf{EEGDeformer}~\citep{ding2024eeg}. The 4 EEG foundation models are \textbf{BIOT}~\citep{yang2024biot}, \textbf{LaBraM}~\citep{jiang2024large}, \textbf{CBraMod}~\citep{wang2024cbramod}, and \textbf{REVE}~\citep{el2026reve}.

\textbf{Shared Training \& Parameter Settings.}
Self-supervised pretraining is conducted for 50 epochs without early stopping, followed by linear probing or fine-tuning for up to 200 epochs with early stopping (patience $=15$) based on the validation F1 score. Batch sizes are set to 512 for pretraining and 128 for supervised learning, linear probing, and fine-tuning. We use AdamW with learning rates of $4\!\times\!10^{-4}$ for pretraining and $1\!\times\!10^{-4}$ for supervised learning, linear probing, and fine-tuning, with a CosineAnnealingLR learning-rate schedule. The patch length for Transformer embedding is set to $L=50$. For self-supervised pretraining, the masking ratio is set to $r=0.5$, with a mixture of random, temporal, and spatial masking strategies. Smooth L1 loss is used as the reconstruction objective with $\beta=0.5$. All subjects in the pretraining datasets are used for self-supervised pretraining. For supervised learning, linear probing, and fine-tuning, we adopt Monte Carlo cross-validation~\citep{wang2025lead} with subject-independent 6:2:2 train/validation/test splits, ensuring strict subject separation while allowing the splits to vary across random seeds. The pretraining is always conducted on single-trial ERP data. Unless otherwise specified, supervised learning, linear probing, and fine-tuning are also conducted on single-trial ERP data; the two case studies in Sections~\ref{sub:trial_averaging_study} and~\ref{sub:subject_level_disease_detection} additionally evaluate averaged-trial ERP data. To distinguish among the model variants evaluated in the following sections, \textbf{we define the model pretrained on 38 single-trial ERP datasets using all predefined settings as the base model of \myname.} For the 4 foundation-model baselines, we fine-tune their officially released pretrained checkpoints, as both the pretraining strategy and pretraining corpus are integral components of a foundation model. We repeat each experiment with 5 random seeds (41-45) and report the mean and standard deviation across runs. We employ 3 evaluation metrics: accuracy, macro-averaged F1 score, and macro-averaged AUROC. All experiments are conducted on 4 NVIDIA RTX A5000 GPUs using Python~3.10 and PyTorch~2.5.1+cu121. Additional details for each method are provided in Appendix~\ref{sec:implementation_details}.

%% file: 060results.tex
\input{tables/results/method_comparison/erp_event_results}

\input{tables/results/method_comparison/erp_disease_results}

\begin{figure*}[h]
    \centering
    \includegraphics[width=1.0\linewidth]{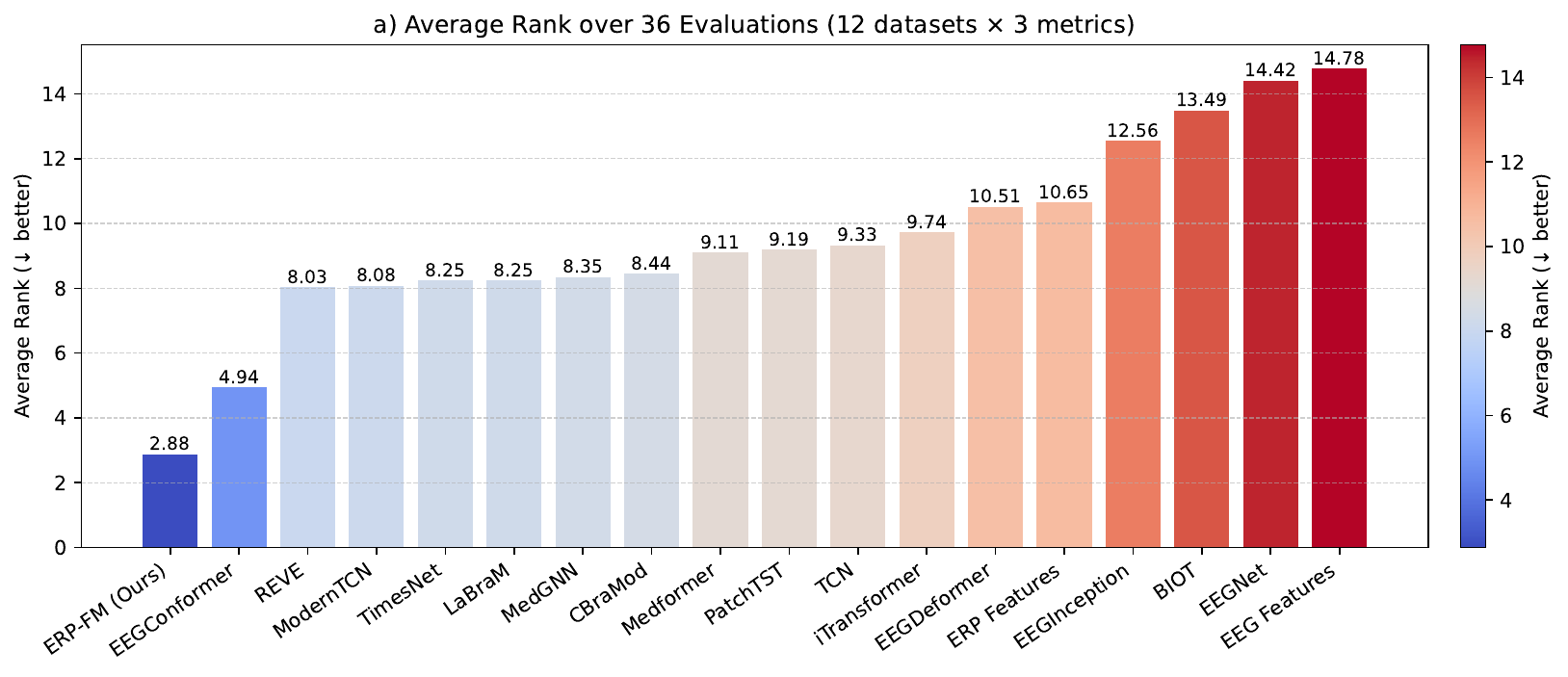}
    \caption{\textbf{Average Performance Rank.} Average performance rank of 18 methods across all 36 evaluations (12 datasets $\times$ 3 metrics). For example, the value 2.88 of \myname indicates an average rank of 2.88. \textbf{Lower ranks and a deeper blue indicate better performance}.
    }
    \label{fig:avg_rank_bar}
    \vspace{-2mm}
\end{figure*}

\subsection{Performance Comparison with Baselines}
\label{sub:performance_comparison}
To train the base model of \myname, we pretrain on 38 ERP datasets and fine-tune the pretrained model on each of the 12 downstream datasets. We compare \myname against 17 baseline methods across these datasets. Accuracy, F1 score, and AUROC results are reported in Tables~\ref{tab:erp_event_results} and~\ref{tab:erp_disease_results}. Overall, \myname achieves Top-1 performance in 20 out of 36 dataset-metric evaluations and obtains the best average ranking across all methods and evaluation settings, as shown in Figure~\ref{fig:avg_rank_bar}. Notably, \myname achieves these results with only approximately 2.4 million parameters and 1.5 million ERP pretraining trials from 3,696 subjects, substantially smaller in model size and pretraining scale than several existing EEG foundation models, including LaBraM, CBraMod, and REVE. For example, REVE contains approximately 69 million parameters and is pretrained on around 60,000 hours of EEG recordings from 25,000 subjects. Despite this large difference in model and pretraining scale, our method still achieves superior overall performance on ERP downstream tasks. Our results are also consistent with the observations reported in ERP-Benchmark~\citep{wang2026benchmarking}, where EEGConformer outperformed existing EEG foundation models and other supervised deep learning methods, achieving the second-best overall ranking in our experiments. Together, these findings suggest that simply increasing model size or pretraining data volume may not be sufficient for ERP representation learning. Instead, the choice of pretraining data, pretraining strategy, and model design may play a more critical role. We further investigate these factors in the following sections.

\input{tables/results/ablation_study/erp_pretrain_study}

\subsection{Effect of ERP Dataset Pretraining Study}
\label{sub:erp_pretrain_study}
To evaluate the effectiveness of self-supervised pretraining on ERP data and the representation-learning capability of \myname, we compare three training strategies using the same backbone architecture: fully supervised learning from scratch, linear probing, and fine-tuning. In the fully supervised setting, the model is randomly initialized and trained directly on each downstream dataset. For linear probing, the pretrained encoder is frozen, and only a linear classifier is optimized for downstream classification. Fine-tuning corresponds to the full \myname model, where the backbone is further optimized on each downstream dataset.

The F1 score results are reported in Table~\ref{tab:erp_pretrain_study}. Fine-tuning achieves the best performance on 8 of the 12 downstream datasets and improves the average F1 score from 62.47\% under supervised training to 65.63\%, with consistent improvements across all 6 ERP cognitive event/condition classification datasets. Linear probing achieves an average F1 score of 63.85\% and substantially outperforms supervised training on several neurological disease detection datasets, including SCPD, RLPD, and AOPD, indicating that the pretrained encoder learns useful transferable representations even without backbone adaptation. PD-SIM and PD-ODD are exceptions where pretraining does not improve performance. Overall, ERP pretraining improves performance across most downstream datasets, although its effectiveness may still vary across datasets.

\input{tables/results/ablation_study/non_erp_pretrain_study}

\begin{figure*}[h]
    \centering
    \includegraphics[width=1.0\linewidth]{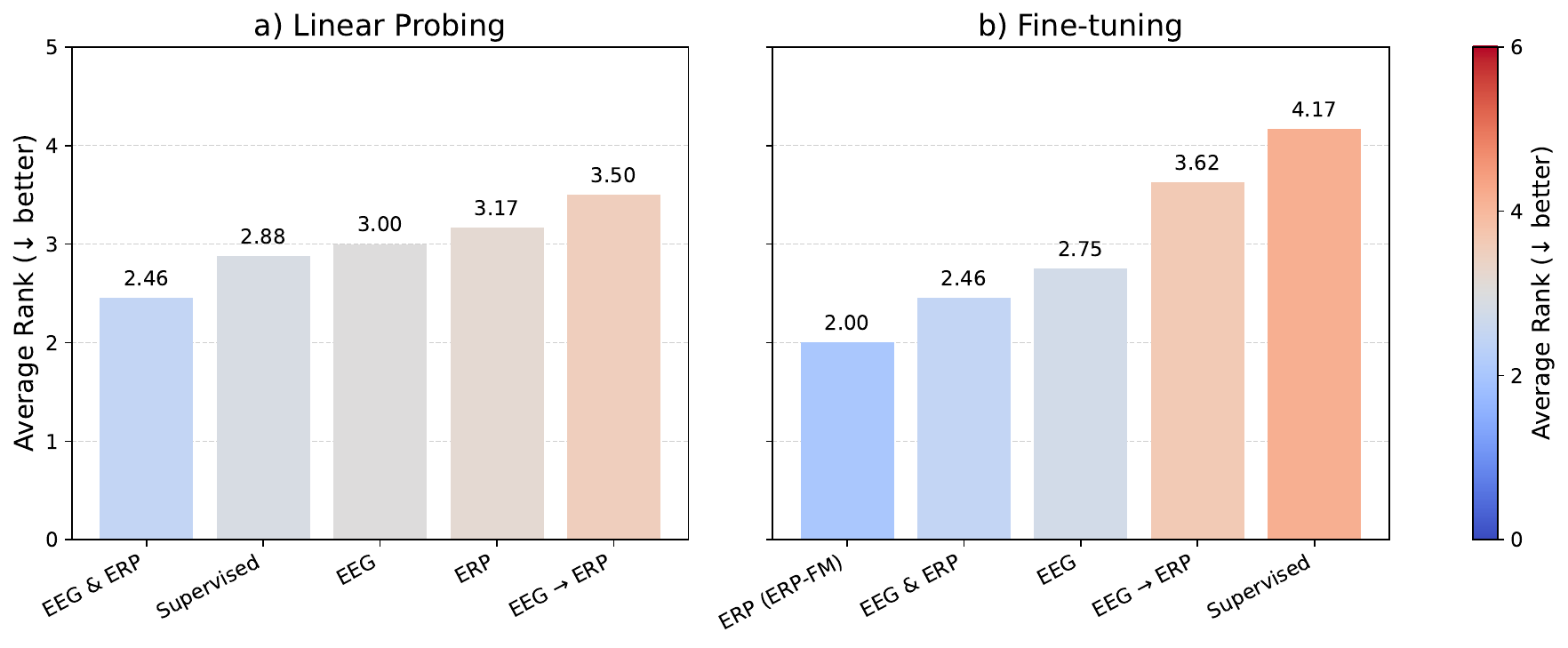}
    \caption{\textbf{Average Performance Rank Across Different Pretraining Resources.} Average performance ranks of supervised training and pretraining with different data resources. Linear probing and fine-tuning settings are ranked separately. \textbf{Lower ranks and deeper blue indicate better performance.}
    }
    \label{fig:non_erp_pretrain_avg_rank_bar}
    \vspace{-3mm}
\end{figure*}

\subsection{Effect of Non-ERP EEG Dataset Pretraining}
\label{sub:non_erp_pretrain_study}
To investigate whether non-ERP EEG data can benefit ERP representation learning, we compare supervised learning with 4 pretraining settings under both linear probing and fine-tuning. \textbf{Supervised} trains the same backbone from scratch without pretraining. \textbf{Non-ERP} and \textbf{ERP} denote pretraining on 5 non-ERP EEG datasets in Table~\ref{tab:non_erp_pretraining_data} and 38 ERP datasets in Table ~\ref{tab:erp_pretraining_data}, respectively. \textbf{Non-ERP \& ERP} jointly pretrains on all 43 datasets, whereas \textbf{Non-ERP $\rightarrow$ ERP} first pretrains on the 5 non-ERP EEG datasets and then performs a second stage of pretraining on the 38 ERP datasets. This sequential strategy is analogous to domain-specialized pretraining in other fields, such as initializing from general-domain image pretraining and then continuing pretraining on domain-specific medical images before downstream adaptation~\citep{yan2025multimodal}. We evaluate all models on the same 12 downstream ERP datasets. 

The results are presented in Tables~\ref{tab:non_erp_pretrain_study_probe} and~\ref{tab:non_erp_pretrain_study_finetune} for linear probing and fine-tuning, respectively, with the average performance ranks shown in Figure~\ref{fig:non_erp_pretrain_avg_rank_bar}. Pretraining on non-ERP EEG alone improves the average F1 score over supervised training from 62.47\% to 63.94\% with linear probing and 64.83\% with fine-tuning. Joint EEG and ERP pretraining further achieves the best average rank under linear probing (2.46) and the second-best rank under fine-tuning (2.46), indicating that non-ERP EEG can be effectively incorporated together with ERP data during pretraining. In contrast, sequential EEG-to-ERP pretraining yields average ranks of 3.50 and 3.62 for linear probing and fine-tuning, respectively, consistently underperforming joint pretraining, Non-ERP-only, and ERP-only pretraining. Notably, ERP-only pretraining remains the strongest strategy under fine-tuning, achieving the best average rank of 2.00. Overall, these results suggest that \textbf{non-ERP EEG contains useful transferable information for ERP downstream tasks}, but its benefit depends on how it is incorporated. Pretraining on non-ERP EEG alone or jointly with ERP is effective, whereas sequential non-ERP-to-ERP pretraining does not provide additional gains over ERP-only or joint pretraining. 

\input{tables/results/ablation_study/patch_length_study}

\subsection{Patch Length Study}
\label{sub:patch_length_study}
Existing EEG foundation models are primarily designed for continuous or non-ERP EEG signals and commonly adopt relatively coarse temporal tokenization. For example, BIOT~\citep{yang2024biot} uses a 1-second temporal window for spectral tokenization, while LaBraM~\citep{jiang2024large} and CBraMod~\citep{wang2024cbramod} use 1-second EEG patches. REVE~\citep{el2026reve} similarly adopts a 1-second patch with a small temporal overlap. Although these temporal resolutions suit longer continuous EEG recordings, they may be suboptimal for ERP signals, whose discriminative information often lies in fine-grained sub-second variations in waveform morphology, latency, and amplitude. To investigate whether patch length affects ERP decoding performance, we vary the patch length $L$ while keeping all other model and training settings unchanged. The full pipeline still pretrains on 38 ERP datasets, then fine-tunes on each of 12 downstream ERP datasets. In addition to $L=50$ used in our base model, corresponding to 0.25 seconds at 200Hz, we evaluate $L=100$ and $L=200$, corresponding to 0.5 and 1.0 seconds, respectively.

The results are presented in Table~\ref{tab:patch_length_study}. The 0.25-second patch achieves the best performance on 8 of the 12 downstream datasets, while the 0.5-second patch performs best on the remaining 4 datasets. In contrast, the 1-second patch consistently achieves the lowest F1 score across all 12 datasets. The average F1 score increases from 63.52\% with a 1-second patch to 65.10\% with a 0.5-second patch and 65.63\% with a 0.25-second patch. These results demonstrate that finer temporal tokenization is particularly important for ERP representation learning, as ERP trials are typically short (e.g., 0.8-2.0 seconds). For a 1-second ERP trial, a non-overlapping 1-second patch produces only one single temporal token per channel, preventing self-attention from explicitly modeling relationships across temporal patches. In contrast, shorter patches better preserve the temporal features of ERP responses and allow the Transformer to capture fine-grained event-related dynamics. Together with the findings in Section~\ref{sub:non_erp_pretrain_study}, which show that pretraining on non-ERP EEG data can still benefit ERP downstream tasks, these results suggest that \textbf{the relatively coarse temporal resolution adopted by existing EEG foundation models may be an important factor limiting their performance on ERP tasks, beyond differences in pretraining data alone.}

\input{tables/results/ablation_study/mask_strategy_study}

\subsection{Masking Strategy Study}
\label{sub:mask_strategy_study}
To investigate the effect of masking strategy, we compare conventional random masking used in the original MAE~\citep{he2022masked} with our mixed masking strategy, which combines random, temporal, and spatial masking during pretraining while keeping all other settings unchanged, as in our base model and fine-tuning.

The results are presented in Table~\ref{tab:mask_ratio_study}. Mixed masking achieves the best performance on 7 of the 12 downstream datasets and improves the average F1 score from 64.45\% to 65.63\%. Although random masking performs better on several individual datasets, its performance is less consistent across datasets, with relatively low F1 scores on PD-SIM (55.75\%) and PD-ODD (54.77\%) but substantially higher performance on SCPD (76.71\%). In comparison, mixed masking provides more consistent overall performance across diverse ERP datasets.

\input{tables/results/case_study/trial_averaging_study}

\subsection{Trial Averaging Study}
\label{sub:trial_averaging_study}
To investigate whether averaged-trial downstream adaptation can complement single-trial ERP pretraining, we compare single-trial and averaged-trial inputs under supervised learning, linear probing, and fine-tuning. Averaged-trial ERPs are constructed following Section~\ref{para:downstream_adaptation}, while the pretrained model remains the same base \myname model pretrained exclusively on single-trial ERP. We exclude datasets without reliable event annotations from this analysis. Specifically, PD-SIM and PD-ODD exhibit unresolved mismatches between trials and available event IDs, while ADHD-WMRI does not provide event IDs in the publicly released data.

The results are presented in Table~\ref{tab:trial_averaging_study}. For the 6 ERP cognitive event/condition classification datasets, the average F1 score across the 6 datasets increases from 62.44\% to 77.60\% for supervised learning, from 61.57\% to 82.67\% for linear probing, and from 64.57\% to 85.08\% for fine-tuning with trial averaging. Notably, even supervised learning without pretraining achieves strong performance after trial averaging, demonstrating the substantial benefit of enhanced ERP signal quality for cognitive event classification. This observation is consistent with conventional ERP analysis, where averaged waveforms under different event or condition types often exhibit visually distinguishable differences in morphology, amplitude, or latency. In contrast, a different situation emerges for neurological disorder classification on SCPD, RLPD, and AOPD. Trial averaging alone provides almost no improvement under supervised learning, with the average F1 score across the 3 datasets increasing only from 60.20\% to 60.59\%, whereas linear probing and fine-tuning improve by 9.01\% and 8.72\%, respectively. Overall, these results indicate that \textbf{trial averaging strategy and ERP pretraining provide complementary benefits.} Trial averaging improves ERP signal quality, while self-supervised pretraining enables more effective downstream learning when the number of averaged trials is limited.

\input{tables/results/case_study/disease_detection_study}

\begin{figure*}[h]
    \centering
    \includegraphics[width=1.0\linewidth]{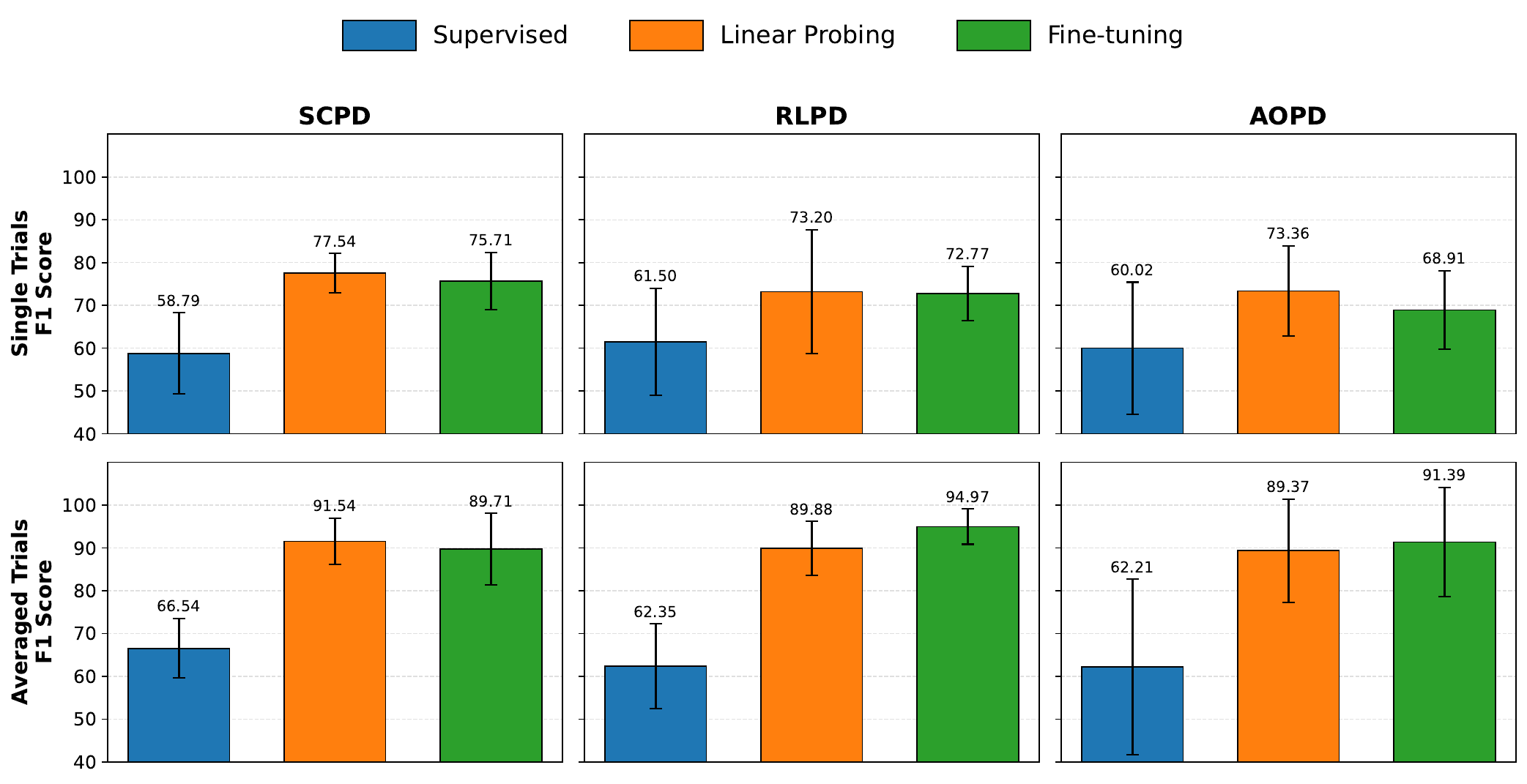}
    \caption{\textbf{Subject-Level Performance Comparison.} F1 score comparison of different downstream adaptation strategies using single or averaged trials. Subject-level results are obtained via majority voting.
    }
    \label{fig:subject_level_performance}
    \vspace{-3mm}
\end{figure*}

\subsection{Subject-Level Neurological Disease Detection}
\label{sub:subject_level_disease_detection}
We further evaluate whether the proposed single-trial pretraining and averaged-trial downstream adaptation framework improves subject-level neurological disease detection. Following Section~\ref{para:downstream_adaptation}, we obtain subject-level predictions through the same majority voting process~\citep{wang2025lead} over all predictions from the same subject. We compare single-trial and averaged-trial downstream adaptation on SCPD, RLPD, and AOPD under supervised learning, linear probing, and fine-tuning.

The results are presented in Table~\ref{tab:disease_detection_study}, with the subject-level F1 score comparison further illustrated in Figure~\ref{fig:subject_level_performance}. For linear probing, the average subject-level F1 score across SCPD, RLPD, and AOPD increases from 74.70\% with single-trial ERP to 90.26\% with averaged-trial ERP, while fine-tuning improves from 72.46\% to 92.02\%. In contrast, supervised learning benefits only marginally from trial averaging, with the average subject-level F1 score increasing from 60.10\% to 63.70\%. Overall, these results identify a highly effective pipeline for ERP-based neurological disease detection: \textbf{single-trial self-supervised pretraining followed by averaged-trial downstream fine-tuning and subject-level majority voting}. Taking RLPD as an example, supervised learning from scratch with single-trial ERP achieves only 60.07\% F1 at the trial-level, which is far from satisfactory for a clinically oriented binary classification task. In contrast, the complete pipeline combining single-trial self-supervised pretraining, averaged-trial fine-tuning, and subject-level majority voting achieves a subject-level F1 score of 94.97\%, representing a substantial improvement over the original 60.07\% trial-level performance. These findings suggest that the proposed pipeline effectively combines the large sample size available during single-trial pretraining with the high signal-to-noise ratio of averaged-trial ERP during downstream adaptation. Moreover, the averaged-trial ERP retains the conventional waveform representation widely used in neuroscience analysis, providing a natural basis for future interpretability studies of disease-related differences in ERP morphology, amplitude, latency, and spatial distribution.

%% file: tables/results/method_comparison/erp_event_results.tex
\begin{table*}[t]
    \centering
    \caption{\textbf{ERP Cognitive Event Classification Results.} Classifying ERP cognitive or behavioral conditions
    associated with ERP-eliciting events, such as target, non-target, and distractor in Oddball. \textcolor{myred}{\textbf{Top-1}}, \textcolor{myblue}{\underline{Top-2}}, and \textcolor{mygreen}{Top-3} results are highlighted in red, blue, and green.
    }
    \vspace{-2mm}
    \label{tab:erp_event_results}
    \resizebox{\textwidth}{!}{
    \begin{tabular}{@{}ll|ccc|ccc|ccc@{}}
    \toprule

    \multicolumn{2}{l|}{\textbf{Datasets}}
    & \multicolumn{3}{c|}{\makecell{\textbf{CESCA-AODD} \\ \textit{(38,151 Trials)} \\ \textit{(127 Subjects, 2 Classes)}} }
    & \multicolumn{3}{c|}{\makecell{\textbf{CESCA-VODD} \\ \textit{(20,419 Trials)} \\ \textit{(127 Subjects, 2 Classes)}} }
    & \multicolumn{3}{c}{\makecell{\textbf{CESCA-FLANKER} \\ \textit{(29,774 Trials)} \\ \textit{(73 Subjects, 2 Classes)}} }
    \\ \midrule

    \multicolumn{2}{l|}{\diagbox{\textbf{Methods}}{\textbf{Metrics}}} & \textbf{Acccuracy} & \textbf{F1 Score} & \textbf{AUROC} & \textbf{Acccuracy} & \textbf{F1 Score} & \textbf{AUROC} & \textbf{Acccuracy} & \textbf{F1 Score} & \textbf{AUROC} \\ \midrule

    \multicolumn{2}{l|}{\textbf{EEG Features}}  & 77.18\std{0.64} & 46.55\std{0.80} & 50.70\std{0.39} & 78.15\std{1.01} & 49.28\std{1.44} & 57.56\std{0.70} & 55.48\std{1.19} & 55.25\std{1.28} & 57.99\std{1.91} \\
    \multicolumn{2}{l|}{\textbf{ERP Features}}  & 75.77\std{4.20} & 47.16\std{2.01} & 51.85\std{0.35} & \textcolor{mygreen}{81.96\std{0.56}} & 64.59\std{0.84} & 77.73\std{1.15} & 64.05\std{1.52} & 63.91\std{1.54} & 69.61\std{1.65} \\
    \midrule
    \multicolumn{2}{l|}{\textbf{TCN}}  & \textcolor{myblue}{77.77\std{1.66}} & 46.71\std{3.14} & 53.35\std{1.94} & \textcolor{myblue}{82.36\std{0.81}} & 66.85\std{1.85} & \textcolor{mygreen}{78.26\std{1.50}} & 63.34\std{1.29} & 63.12\std{1.33} & 68.83\std{1.54} \\
    \multicolumn{2}{l|}{\textbf{ModernTCN}}  & 75.39\std{1.19} & 53.54\std{0.57} & 60.56\std{1.30} & 81.81\std{0.78} & 66.46\std{2.81} & 78.15\std{1.13} & 64.16\std{0.61} & 64.09\std{0.64} & 70.00\std{1.19} \\
    \multicolumn{2}{l|}{\textbf{TimesNet}}  & 75.49\std{1.42} & 53.81\std{2.03} & 61.18\std{0.85} & 81.28\std{0.72} & 67.67\std{1.59} & 77.57\std{1.16} & 63.02\std{0.83} & 62.89\std{0.83} & 68.45\std{1.15} \\
    \multicolumn{2}{l|}{\textbf{PatchTST}}  & 76.86\std{0.67} & 53.25\std{1.02} & \textcolor{mygreen}{62.31\std{1.32}} & 81.02\std{0.86} & \textcolor{mygreen}{67.90\std{1.69}} & 77.96\std{0.97} & 63.76\std{0.78} & 63.68\std{0.79} & 69.23\std{1.14} \\
    \multicolumn{2}{l|}{\textbf{iTransformer}}  & 75.78\std{1.06} & 53.66\std{0.70} & 61.10\std{1.42} & 81.19\std{0.65} & 65.77\std{1.43} & 76.27\std{1.21} & 63.67\std{1.27} & 63.64\std{1.26} & 69.46\std{1.59} \\
    \multicolumn{2}{l|}{\textbf{Medformer}}  & 74.17\std{1.09} & \textcolor{myred}{\textbf{56.63\std{0.80}}} & \textcolor{myred}{\textbf{62.58\std{1.33}}} & 80.52\std{0.74} & 66.95\std{1.30} & 76.00\std{1.74} & 63.57\std{0.59} & 63.51\std{0.60} & 69.26\std{1.12} \\
    \multicolumn{2}{l|}{\textbf{MedGNN}}  & 75.18\std{0.93} & 54.36\std{0.90} & 61.42\std{1.34} & 81.50\std{1.03} & 67.30\std{1.75} & 77.26\std{1.79} & 64.03\std{1.00} & 63.94\std{1.01} & \textcolor{mygreen}{70.22\std{1.57}} \\
    \multicolumn{2}{l|}{\textbf{EEGNet}}  & \textcolor{myred}{\textbf{79.10\std{0.02}}} & 44.16\std{0.01} & 50.17\std{0.45} & 77.85\std{2.21} & 47.84\std{3.25} & 57.24\std{1.65} & 51.60\std{0.69} & 51.25\std{0.79} & 52.70\std{0.80} \\
    \multicolumn{2}{l|}{\textbf{EEGInception}}  & 73.21\std{10.82} & 45.02\std{1.04} & 51.07\std{0.90} & 73.76\std{2.54} & 61.75\std{1.59} & 70.55\std{2.36} & 59.12\std{1.17} & 58.47\std{1.50} & 63.13\std{1.70} \\
    \multicolumn{2}{l|}{\textbf{EEGConformer}}  & 74.39\std{0.62} & 54.48\std{0.76} & 60.63\std{1.13} & 81.33\std{0.53} & \textcolor{myblue}{69.17\std{1.18}} & \textcolor{myblue}{79.45\std{1.14}} & \textcolor{mygreen}{64.24\std{0.72}} & \textcolor{mygreen}{64.11\std{0.79}} & 70.00\std{1.03} \\
    \multicolumn{2}{l|}{\textbf{EEGDeformer}}  & 75.05\std{3.29} & 54.36\std{4.41} & 61.71\std{3.97} & 80.37\std{0.74} & 49.61\std{2.36} & 71.47\std{6.39} & 63.63\std{2.72} & 63.45\std{2.88} & 69.18\std{3.64} \\
    \midrule
    \multicolumn{2}{l|}{\textbf{BIOT}}  & \textcolor{mygreen}{77.29\std{3.62}} & 45.04\std{1.76} & 49.98\std{0.44} & 73.11\std{2.65} & 54.15\std{1.19} & 58.78\std{0.75} & 54.51\std{1.11} & 54.11\std{1.03} & 56.21\std{1.32} \\
    \multicolumn{2}{l|}{\textbf{LaBraM}}  & 74.47\std{1.19} & \textcolor{myblue}{55.60\std{0.69}} & 61.79\std{1.32} & 80.10\std{0.77} & 65.45\std{1.28} & 75.28\std{1.35} & 63.51\std{1.15} & 63.41\std{1.12} & 68.94\std{1.77} \\
    \multicolumn{2}{l|}{\textbf{CBraMod}}  & 76.30\std{1.40} & 52.98\std{1.33} & 60.87\std{1.97} & 81.46\std{1.16} & 66.30\std{1.66} & 76.65\std{1.49} & 64.11\std{1.22} & 64.00\std{1.28} & 70.03\std{1.37} \\
    \multicolumn{2}{l|}{\textbf{REVE}}  & 74.06\std{2.19} & \textcolor{mygreen}{55.40\std{1.16}} & \textcolor{myblue}{62.52\std{0.97}} & 81.03\std{1.94} & 67.77\std{3.29} & 76.88\std{3.24} & \textcolor{myblue}{65.32\std{1.52}} & \textcolor{myblue}{65.20\std{1.58}} & \textcolor{myblue}{71.06\std{1.91}} \\
    \midrule

    \multicolumn{2}{l|}{\textbf{ERP-FM (Ours)}}  & 75.74\std{0.75} & 54.61\std{1.09} & 62.15\std{1.30} & \textcolor{myred}{\textbf{82.46\std{0.59}}} & \textcolor{myred}{\textbf{69.75\std{1.44}}} & \textcolor{myred}{\textbf{80.33\std{1.32}}} & \textcolor{myred}{\textbf{65.48\std{1.13}}} & \textcolor{myred}{\textbf{65.44\std{1.11}}} & \textcolor{myred}{\textbf{71.27\std{1.80}}}  \\
    \midrule

    \midrule
    \multicolumn{2}{l|}{\textbf{Datasets}}
    & \multicolumn{3}{c|}{\makecell{\textbf{TDBrain-ODD} \\ \textit{(54,400 Trials)} \\ \textit{(127 Subjects, 2 Classes)}} }
    & \multicolumn{3}{c|}{\makecell{\textbf{NSERP-MSIT} \\ \textit{(16,729 Trials)} \\ \textit{(42 Subjects, 4 Classes)}} }
    & \multicolumn{3}{c}{\makecell{\textbf{NSERP-ODD}  \\ \textit{(27,865 Trials)} \\ \textit{(42 Subjects, 3 Classes)}} }
    \\ \midrule

    \multicolumn{2}{l|}{\diagbox{\textbf{Methods}}{\textbf{Metrics}}} & \textbf{Acccuracy} & \textbf{F1 Score} & \textbf{AUROC} & \textbf{Acccuracy} & \textbf{F1 Score} & \textbf{AUROC} & \textbf{Acccuracy} & \textbf{F1 Score} & \textbf{AUROC} \\ \midrule

    \multicolumn{2}{l|}{\textbf{EEG Features}}  & 86.22\std{0.35} & 71.43\std{1.79} & 85.38\std{0.43} & 26.34\std{0.78} & 25.81\std{0.54} & 52.16\std{0.69} & 74.01\std{1.10} & 38.98\std{2.34} & 68.63\std{2.79} \\
    \multicolumn{2}{l|}{\textbf{ERP Features}}  & 92.03\std{0.52} & 85.20\std{1.29} & 94.83\std{0.94} & 35.96\std{2.13} & 35.23\std{2.09} & 63.41\std{2.16} & 81.74\std{1.81} & 61.15\std{3.73} & 87.47\std{2.26} \\
    \midrule
    \multicolumn{2}{l|}{\textbf{TCN}}  & \textcolor{mygreen}{93.42\std{0.34}} & 88.24\std{0.75} & 95.99\std{0.67} & 35.56\std{2.32} & 34.91\std{2.28} & 62.81\std{2.47} & 84.19\std{1.40} & \textcolor{mygreen}{66.77\std{3.27}} & 90.33\std{1.84} \\
    \multicolumn{2}{l|}{\textbf{ModernTCN}}  & 92.43\std{0.42} & 86.34\std{0.88} & 94.69\std{1.00} & 37.16\std{2.49} & 36.80\std{2.35} & 65.03\std{2.87} & 83.01\std{1.07} & 62.76\std{2.70} & 88.75\std{1.37} \\
    \multicolumn{2}{l|}{\textbf{TimesNet}}  & 91.99\std{0.33} & 86.34\std{0.72} & 95.18\std{0.71} & 37.05\std{2.51} & 36.53\std{2.44} & 65.28\std{2.82} & 82.88\std{1.06} & 64.22\std{2.77} & 88.84\std{1.72} \\
    \multicolumn{2}{l|}{\textbf{PatchTST}}  & 92.26\std{0.36} & 86.57\std{1.09} & 94.94\std{1.08} & 36.34\std{2.28} & 35.95\std{2.23} & 63.97\std{2.62} & 82.72\std{1.08} & 63.54\std{2.93} & 88.29\std{1.65} \\
    \multicolumn{2}{l|}{\textbf{iTransformer}}  & 91.37\std{0.60} & 85.19\std{1.08} & 94.00\std{1.12} & 36.62\std{2.38} & 36.28\std{2.30} & 64.89\std{2.63} & 81.46\std{1.24} & 61.56\std{2.57} & 87.47\std{1.84} \\
    \multicolumn{2}{l|}{\textbf{Medformer}}  & 91.77\std{0.41} & 85.49\std{0.96} & 93.90\std{0.99} & 37.99\std{2.79} & 37.38\std{2.52} & 66.55\std{3.06} & 81.29\std{1.08} & 63.38\std{2.50} & 87.97\std{1.55} \\
    \multicolumn{2}{l|}{\textbf{MedGNN}}  & 93.13\std{0.36} & 87.80\std{0.80} & 95.77\std{0.69} & 38.44\std{2.40} & 37.66\std{2.46} & 66.79\std{2.63} & 83.77\std{1.17} & 65.30\std{2.53} & 90.10\std{1.61} \\
    \multicolumn{2}{l|}{\textbf{EEGNet}}  & 74.10\std{3.64} & 59.09\std{1.58} & 69.65\std{1.38} & 25.72\std{1.36} & 19.86\std{2.29} & 49.92\std{1.09} & 62.34\std{8.41} & 36.75\std{2.62} & 61.87\std{1.52} \\
    \multicolumn{2}{l|}{\textbf{EEGInception}}  & 90.93\std{0.35} & 83.51\std{0.74} & 92.75\std{0.56} & 31.79\std{1.87} & 27.27\std{3.63} & 59.97\std{2.86} & 77.56\std{2.29} & 58.46\std{2.65} & 81.95\std{2.67} \\
    \multicolumn{2}{l|}{\textbf{EEGConformer}}  & 93.38\std{0.21} & \textcolor{mygreen}{88.72\std{0.37}} & \textcolor{myblue}{96.44\std{0.65}} & \textcolor{myblue}{38.99\std{2.14}} & \textcolor{mygreen}{38.49\std{2.12}} & \textcolor{mygreen}{67.14\std{2.36}} & \textcolor{mygreen}{84.28\std{1.27}} & \textcolor{myblue}{68.02\std{2.54}} & \textcolor{mygreen}{90.56\std{1.48}} \\
    \multicolumn{2}{l|}{\textbf{EEGDeformer}}  & 89.55\std{1.30} & 78.28\std{3.90} & 91.30\std{2.01} & \textcolor{myred}{\textbf{39.28\std{2.80}}} & \textcolor{myred}{\textbf{38.89\std{2.81}}} & \textcolor{myred}{\textbf{67.58\std{2.94}}} & \textcolor{myblue}{84.84\std{1.73}} & 66.50\std{5.13} & \textcolor{myred}{\textbf{91.96\std{1.49}}} \\
    \midrule
    \multicolumn{2}{l|}{\textbf{BIOT}}  & 82.85\std{1.33} & 69.01\std{0.84} & 80.07\std{1.26} & 29.36\std{0.71} & 28.90\std{0.73} & 56.02\std{0.93} & 75.78\std{1.15} & 49.37\std{2.07} & 74.69\std{2.51} \\
    \multicolumn{2}{l|}{\textbf{LaBraM}}  & 92.53\std{0.53} & 86.62\std{1.31} & 94.79\std{0.89} & 36.15\std{2.50} & 35.79\std{2.78} & 64.44\std{2.51} & 82.45\std{1.35} & 64.28\std{2.65} & 88.44\std{1.79} \\
    \multicolumn{2}{l|}{\textbf{CBraMod}}  & 92.89\std{0.28} & 86.99\std{0.84} & 95.10\std{0.62} & 38.54\std{1.87} & 38.01\std{2.00} & 66.30\std{2.41} & 84.02\std{1.12} & 66.51\std{2.70} & 89.77\std{1.70} \\
    \multicolumn{2}{l|}{\textbf{REVE}}  & \textcolor{myblue}{93.73\std{1.36}} & \textcolor{myblue}{88.74\std{2.40}} & \textcolor{mygreen}{96.06\std{1.73}} & 35.04\std{1.69} & 34.73\std{1.81} & 63.32\std{2.09} & 81.71\std{1.73} & 61.42\std{4.59} & 87.43\std{2.44} \\

    \midrule
    \multicolumn{2}{l|}{\textbf{ERP-FM (Ours)}}  & \textcolor{myred}{\textbf{94.34\std{0.57}}} & \textcolor{myred}{\textbf{90.21\std{0.95}}} & \textcolor{myred}{\textbf{97.15\std{0.73}}} & \textcolor{mygreen}{38.82\std{2.61}} & \textcolor{myblue}{38.49\std{2.52}} & \textcolor{myblue}{67.34\std{3.04}} & \textcolor{myred}{\textbf{84.99\std{1.12}}} & \textcolor{myred}{\textbf{68.93\std{2.49}}} & \textcolor{myblue}{90.95\std{1.77}}  \\

    \bottomrule
    \end{tabular}
    }
\vspace{-3mm}
\end{table*}

%% file: tables/results/method_comparison/erp_disease_results.tex
\begin{table*}[t]
    \centering
    \caption{\textbf{ERP-Based Neurological Disease Detection Results.} ERP-based neurological disease detection focuses on classifying neurological disorders, such as PD or ADHD. \textcolor{myred}{\textbf{Top-1}}, \textcolor{myblue}{\underline{Top-2}}, and \textcolor{mygreen}{Top-3} results are highlighted in red, blue, and green. 
    }
    \vspace{-2mm}
    \label{tab:erp_disease_results}
    \resizebox{\textwidth}{!}{
    \begin{tabular}{@{}ll|ccc|ccc|ccc@{}}
    \toprule

    \multicolumn{2}{l|}{\textbf{Datasets}}
    & \multicolumn{3}{c|}{ \makecell{\textbf{PD-SIM} \\ \textit{(55,921 Trials)} \\ \textit{(147 Subjects, 2 Classes)}} }
    & \multicolumn{3}{c|}{\makecell{\textbf{PD-ODD} \\ \textit{(34,702 Trials)} \\ \textit{(146 Subjects, 2 Classes)}} }
    & \multicolumn{3}{c}{\makecell{\textbf{ADHD-WMRI} \\ \textit{(21,832 Trials)} \\ \textit{(59 Subjects, 2 Classes)}}  }
    \\ \midrule

    \multicolumn{2}{l|}{\diagbox{\textbf{Methods}}{\textbf{Metrics}}} & \textbf{Acccuracy} & \textbf{F1 Score} & \textbf{AUROC} & \textbf{Acccuracy} & \textbf{F1 Score} & \textbf{AUROC} & \textbf{Acccuracy} & \textbf{F1 Score} & \textbf{AUROC} \\ \midrule

    \multicolumn{2}{l|}{\textbf{EEG Features}}  & 67.68\std{2.04} & 61.08\std{2.95} & 70.49\std{3.07} & 66.29\std{2.70} & 59.80\std{4.47} & 68.83\std{4.30} & 59.33\std{3.11} & 56.82\std{4.10} & 58.40\std{6.85}  \\
    \multicolumn{2}{l|}{\textbf{ERP Features}}  & 67.32\std{2.33} & 61.20\std{3.14} & 71.20\std{2.43} & 68.93\std{2.13} & 64.37\std{2.66} & 73.40\std{2.73} & 59.17\std{3.52} & 57.14\std{4.29} & 59.28\std{7.88}  \\

    \midrule
    \multicolumn{2}{l|}{\textbf{TCN}}  & 62.86\std{4.08} & 55.55\std{5.90} & 62.67\std{6.45} & 66.71\std{2.54} & 59.64\std{3.87} & 69.02\std{2.56} & 62.92\std{2.34} & 60.80\std{1.98} & 68.08\std{3.22}  \\
    \multicolumn{2}{l|}{\textbf{ModernTCN}}  & \textcolor{myred}{\textbf{72.35\std{2.11}}} & \textcolor{myblue}{66.49\std{2.07}} & \textcolor{myred}{\textbf{76.44\std{3.27}}} & \textcolor{myred}{\textbf{72.99\std{2.52}}} & \textcolor{myred}{\textbf{67.52\std{1.35}}} & \textcolor{myred}{\textbf{77.55\std{2.39}}} & 64.47\std{2.13} & 61.42\std{3.47} & 68.77\std{4.11}  \\
    \multicolumn{2}{l|}{\textbf{TimesNet}}  & 65.71\std{4.46} & 55.04\std{6.98} & 63.80\std{10.08} & 68.62\std{2.97} & 58.91\std{4.28} & 68.56\std{4.60} & 66.95\std{3.57} & \textcolor{mygreen}{64.97\std{3.80}} & \textcolor{myblue}{74.22\std{4.77}}  \\
    \multicolumn{2}{l|}{\textbf{PatchTST}}  & \textcolor{mygreen}{70.34\std{1.81}} & \textcolor{mygreen}{65.71\std{3.71}} & \textcolor{mygreen}{74.59\std{4.70}} & \textcolor{myblue}{71.35\std{3.45}} & \textcolor{myblue}{66.75\std{2.25}} & \textcolor{myblue}{76.76\std{3.11}} & 64.12\std{2.28} & 62.38\std{2.27} & 69.53\std{3.03}  \\
    \multicolumn{2}{l|}{\textbf{iTransformer}}  & \textcolor{myblue}{71.54\std{2.31}} & \textcolor{myred}{\textbf{66.98\std{2.87}}} & \textcolor{myblue}{76.08\std{2.89}} & \textcolor{mygreen}{70.89\std{3.67}} & \textcolor{mygreen}{66.47\std{3.26}} & \textcolor{mygreen}{76.11\std{3.86}} & 64.57\std{2.29} & 62.11\std{3.19} & 69.73\std{3.59}  \\
    \multicolumn{2}{l|}{\textbf{Medformer}}  & 68.11\std{1.04} & 61.69\std{3.59} & 70.55\std{2.74} & 68.78\std{3.32} & 62.66\std{1.93} & 72.95\std{3.31} & 66.29\std{2.46} & 63.45\std{3.75} & 71.18\std{3.89}  \\
    \multicolumn{2}{l|}{\textbf{MedGNN}}  & 60.13\std{7.44} & 53.84\std{7.04} & 62.22\std{9.83} & 67.44\std{2.04} & 62.34\std{3.83} & 72.40\std{5.75} & 63.95\std{5.82} & 60.60\std{7.25} & 67.74\std{9.00}  \\
    \multicolumn{2}{l|}{\textbf{EEGNet}}  & 67.56\std{1.94} & 64.63\std{1.62} & 72.01\std{2.77} & 62.62\std{1.19} & 59.97\std{1.36} & 68.76\std{2.61} & 51.86\std{6.57} & 45.70\std{3.67} & 44.28\std{5.15}  \\
    \multicolumn{2}{l|}{\textbf{EEGInception}}  & 62.53\std{6.00} & 53.60\std{8.61} & 58.11\std{10.20} & 63.62\std{5.70} & 59.61\std{5.38} & 67.44\std{6.81} & 66.27\std{3.91} & 64.01\std{4.35} & 71.72\std{5.37}  \\
    \multicolumn{2}{l|}{\textbf{EEGConformer}}  & 61.32\std{6.19} & 56.03\std{6.24} & 60.74\std{7.89} & 68.89\std{2.25} & 63.19\std{3.66} & 73.71\std{3.68} & \textcolor{myred}{\textbf{70.22\std{4.06}}} & \textcolor{myred}{\textbf{67.81\std{5.95}}} & \textcolor{myred}{\textbf{76.69\std{5.95}}}  \\
    \multicolumn{2}{l|}{\textbf{EEGDeformer}}  & 61.95\std{4.83} & 55.17\std{5.57} & 62.00\std{7.30} & 68.00\std{2.74} & 63.25\std{3.70} & 73.95\std{4.00} & 60.00\std{1.43} & 57.19\std{1.77} & 61.01\std{2.72}  \\

    \midrule
    \multicolumn{2}{l|}{\textbf{BIOT}}  & 63.48\std{3.17} & 56.83\std{5.22} & 63.03\std{7.93} & 65.63\std{5.08} & 60.27\std{6.14} & 69.41\std{6.77} & 55.04\std{7.07} & 53.87\std{6.79} & 56.14\std{9.22}  \\
    \multicolumn{2}{l|}{\textbf{LaBraM}}  & 62.38\std{3.62} & 55.32\std{5.58} & 62.56\std{5.71} & 69.61\std{4.87} & 64.63\std{6.75} & 72.14\std{6.10} & 66.32\std{9.79} & 64.57\std{9.70} & 71.91\std{12.90}  \\
    \multicolumn{2}{l|}{\textbf{CBraMod}}  & 66.34\std{2.43} & 59.56\std{1.70} & 67.42\std{3.39} & 67.95\std{5.58} & 63.66\std{6.44} & 72.97\std{7.61} & 64.44\std{4.58} & 62.72\std{4.79} & 69.05\std{6.38}  \\
    \multicolumn{2}{l|}{\textbf{REVE}}  & 63.08\std{4.22} & 54.74\std{7.65} & 62.30\std{13.18} & 65.12\std{10.05} & 57.94\std{9.30} & 68.65\std{12.38} & \textcolor{mygreen}{67.41\std{4.64}} & 64.69\std{5.01} & 72.78\std{7.08}  \\

    \midrule
    \multicolumn{2}{l|}{\textbf{ERP-FM (Ours)}}  & 67.94\std{5.72} & 61.12\std{6.46} & 68.23\std{11.42} & 68.81\std{3.89} & 60.22\std{4.51} & 70.47\std{5.87} & \textcolor{myblue}{68.38\std{3.47}} & \textcolor{myblue}{65.93\std{4.36}} & \textcolor{mygreen}{74.06\std{6.63}}  \\
    \midrule

    \midrule
    \multicolumn{2}{l|}{\textbf{Datasets}}
    & \multicolumn{3}{c|}{\makecell{\textbf{SCPD} \\ \textit{(46,193 Trials)} \\ \textit{(55 Subjects, 2 Classes)}} }
    & \multicolumn{3}{c|}{\makecell{\textbf{RLPD} \\ \textit{(21,510 Trials)} \\ \textit{(56 Subjects, 2 Classes)}} }
    & \multicolumn{3}{c}{\makecell{\textbf{AOPD}  \\ \textit{(14,625 Trials)} \\ \textit{(50 Subjects, 2 Classes)}} }
    \\ \midrule

    \multicolumn{2}{l|}{\diagbox{\textbf{Methods}}{\textbf{Metrics}}} & \textbf{Acccuracy} & \textbf{F1 Score} & \textbf{AUROC} & \textbf{Acccuracy} & \textbf{F1 Score} & \textbf{AUROC} & \textbf{Acccuracy} & \textbf{F1 Score} & \textbf{AUROC} \\ \midrule

    \multicolumn{2}{l|}{\textbf{EEG Features}}  & 66.31\std{4.14} & 62.18\std{3.91} & 70.19\std{4.80} & 62.42\std{3.11} & 56.88\std{4.11} & 62.37\std{5.26} & 63.05\std{3.51} & 58.93\std{2.35} & 63.85\std{4.03}  \\
    \multicolumn{2}{l|}{\textbf{ERP Features}}  & 67.02\std{5.01} & 62.35\std{5.09} & 69.61\std{7.11} & 64.96\std{3.03} & 61.23\std{3.61} & 67.79\std{4.91} & 63.71\std{2.51} & 59.04\std{2.00} & 64.09\std{1.62}  \\

    \midrule
    \multicolumn{2}{l|}{\textbf{TCN}}  & 67.55\std{3.74} & 63.52\std{5.76} & 71.58\std{8.21} & 70.07\std{3.76} & 65.21\std{3.10} & 74.65\std{5.16} & 70.66\std{10.64} & \textcolor{mygreen}{67.48\std{10.64}} & 74.54\std{12.71}  \\
    \multicolumn{2}{l|}{\textbf{ModernTCN}}  & 68.84\std{6.27} & 64.58\std{5.94} & 73.07\std{7.20} & 66.13\std{4.15} & 59.68\std{3.84} & 67.19\std{6.41} & 62.05\std{11.07} & 53.80\std{6.58} & 58.80\std{5.91}  \\
    \multicolumn{2}{l|}{\textbf{TimesNet}}  & \textcolor{myblue}{70.75\std{2.85}} & 64.50\std{3.12} & 73.63\std{3.84} & 71.95\std{5.51} & 67.27\std{5.00} & 75.64\std{7.54} & 71.28\std{7.74} & 65.58\std{8.43} & 73.99\std{11.50}  \\
    \multicolumn{2}{l|}{\textbf{PatchTST}}  & 66.63\std{6.02} & 62.32\std{5.21} & 71.08\std{7.80} & 64.86\std{3.93} & 60.65\std{4.83} & 67.05\std{5.81} & 59.81\std{7.17} & 56.58\std{6.14} & 60.37\std{7.82}  \\
    \multicolumn{2}{l|}{\textbf{iTransformer}}  & 68.26\std{5.37} & 64.30\std{4.49} & 72.68\std{6.31} & 64.66\std{3.59} & 59.48\std{4.14} & 66.24\std{6.13} & 60.60\std{9.49} & 57.05\std{7.63} & 60.43\std{10.54}  \\
    \multicolumn{2}{l|}{\textbf{Medformer}}  & 67.73\std{4.41} & 62.06\std{5.13} & 69.14\std{7.44} & 67.96\std{3.07} & 63.07\std{4.53} & 70.25\std{5.47} & 68.15\std{8.86} & 63.44\std{8.36} & 70.66\std{11.09}  \\
    \multicolumn{2}{l|}{\textbf{MedGNN}}  & 66.55\std{6.18} & 61.52\std{6.68} & 72.49\std{9.47} & 72.25\std{6.26} & 66.93\std{6.80} & 75.16\std{7.76} & \textcolor{myblue}{74.25\std{6.25}} & 67.46\std{5.84} & 73.07\std{6.47}  \\
    \multicolumn{2}{l|}{\textbf{EEGNet}}  & 67.84\std{6.46} & 64.97\std{4.89} & \textcolor{mygreen}{74.74\std{6.46}} & 64.52\std{3.89} & 61.86\std{2.95} & 69.34\std{5.12} & 62.19\std{9.58} & 59.72\std{7.97} & 67.66\std{7.09}  \\
    \multicolumn{2}{l|}{\textbf{EEGInception}}  & 68.21\std{6.32} & 65.01\std{6.20} & 72.80\std{8.25} & 71.72\std{4.22} & 68.22\std{4.35} & 76.72\std{5.21} & 66.21\std{9.78} & 64.20\std{10.04} & 75.07\std{13.57}  \\
    \multicolumn{2}{l|}{\textbf{EEGConformer}}  & 69.57\std{5.38} & 65.66\std{6.06} & 72.75\std{7.98} & \textcolor{mygreen}{72.47\std{4.84}} & \textcolor{mygreen}{69.16\std{4.68}} & \textcolor{myblue}{78.48\std{5.52}} & \textcolor{mygreen}{71.87\std{8.70}} & \textcolor{myblue}{68.49\std{8.28}} & \textcolor{myblue}{77.57\std{11.67}}  \\
    \multicolumn{2}{l|}{\textbf{EEGDeformer}}  & 63.41\std{14.05} & 59.64\std{13.25} & 72.63\std{11.38} & 66.03\std{6.83} & 62.21\std{5.23} & 70.98\std{6.41} & 65.94\std{6.92} & 62.27\std{5.38} & 68.74\std{7.33}  \\

    \midrule
    \multicolumn{2}{l|}{\textbf{BIOT}}  & \textcolor{mygreen}{69.74\std{3.86}} & \textcolor{myblue}{66.45\std{3.69}} & 73.63\std{4.51} & 70.22\std{4.68} & 66.23\std{5.80} & 74.83\std{5.87} & 61.17\std{8.22} & 57.90\std{7.43} & 62.16\std{10.35}  \\
    \multicolumn{2}{l|}{\textbf{LaBraM}}  & 68.47\std{5.43} & \textcolor{mygreen}{66.27\std{5.56}} & \textcolor{myblue}{75.15\std{8.27}} & 72.04\std{2.31} & 67.89\std{2.96} & \textcolor{mygreen}{77.62\std{4.59}} & 68.61\std{12.66} & 66.33\std{11.97} & 69.94\std{14.04}  \\
    \multicolumn{2}{l|}{\textbf{CBraMod}}  & 63.53\std{3.88} & 60.15\std{2.90} & 68.82\std{6.37} & 69.09\std{4.21} & 66.38\std{3.88} & 76.19\std{3.90} & 63.80\std{11.82} & 61.36\std{10.45} & 68.47\std{13.49}  \\
    \multicolumn{2}{l|}{\textbf{REVE}}  & 69.69\std{5.65} & 64.22\std{5.89} & 72.90\std{4.85} & \textcolor{myblue}{74.75\std{6.35}} & \textcolor{myblue}{71.04\std{6.91}} & 76.48\std{10.99} & 67.61\std{9.82} & 65.12\std{9.06} & \textcolor{mygreen}{75.13\std{8.94}}  \\

    \midrule
    \multicolumn{2}{l|}{\textbf{ERP-FM (Ours)}}  & \textcolor{myred}{\textbf{76.65\std{4.76}}} & \textcolor{myred}{\textbf{71.71\std{4.97}}} & \textcolor{myred}{\textbf{81.04\std{4.71}}} & \textcolor{myred}{\textbf{74.81\std{2.96}}} & \textcolor{myred}{\textbf{71.42\std{3.22}}} & \textcolor{myred}{\textbf{79.55\std{5.76}}} & \textcolor{myred}{\textbf{74.32\std{4.91}}} & \textcolor{myred}{\textbf{69.75\std{5.16}}} & \textcolor{myred}{\textbf{79.27\std{8.35}}}  \\

    \bottomrule
    \end{tabular}
    }
\vspace{-3mm}
\end{table*}

%% file: tables/results/ablation_study/erp_pretrain_study.tex
\begin{table*}[t]
    \centering
    \caption{\textbf{Effect of ERP Pretraining.} F1 score comparison of supervised training from scratch, linear probing, and fine-tuning using the same \myname backbone across 12 downstream ERP datasets.
    }
    \vspace{-2mm}
    \label{tab:erp_pretrain_study}
    \resizebox{1.0\textwidth}{!}{%
    \begin{tabular}{@{}l|c|c|c|c|c|c@{}}
    \toprule
    \multicolumn{1}{l|}{\diagbox{\textbf{Models}}{\textbf{Datasets}}}  & \textbf{CESCA-AODD} & \textbf{CESCA-VODD} &  \textbf{CESCA-FLANKER} &  \textbf{TDBrain-ODD}  & \textbf{NSERP-MSIT} & \textbf{NSERP-ODD} \\
    \midrule
    \multicolumn{1}{l|}{\textbf{Supervised}} & 53.90\std{0.86} & 67.39\std{1.98} & 64.03\std{1.15} & 86.67\std{0.83} & 37.79\std{2.25} & 64.88\std{2.27} \\
    \multicolumn{1}{l|}{\textbf{Linear Probing}} & 52.30\std{0.67} & 65.00\std{2.12} & 64.03\std{0.55} & 86.79\std{1.26} & 36.26\std{2.01} & 65.06\std{2.13} \\
    \multicolumn{1}{l|}{\textbf{Fine-tuning (ERP-FM)}}  & \textbf{54.61\std{1.09}} & \textbf{69.75\std{1.44}} & \textbf{65.44\std{1.11}} & \textbf{90.21\std{0.95}} & \textbf{38.49\std{2.52}} & \textbf{68.93\std{2.49}} \\
    
    \midrule

    \midrule
    \multicolumn{1}{l|}{\diagbox{\textbf{Models}}{\textbf{Datasets}}}  & \textbf{PD-SIM}  & \textbf{PD-ODD}  & \textbf{ADHD-WMRI}  & \textbf{SCPD}  & \textbf{RLPD}  & \textbf{AOPD}  \\
    \midrule
    \multicolumn{1}{l|}{\textbf{Supervised}} & \textbf{65.20\std{3.74}} & \textbf{68.15\std{3.06}} & 60.97\std{3.88} & 63.25\std{5.25} & 60.07\std{5.40} & 57.29\std{7.37} \\
    \multicolumn{1}{l|}{\textbf{Linear Probing}} & 57.26\std{7.83} & 62.10\std{6.43} & 61.60\std{3.11} & \textbf{72.83\std{3.23}} & 71.06\std{8.11} & \textbf{71.89\std{3.51}} \\
    \multicolumn{1}{l|}{\textbf{Fine-tuning (ERP-FM)}}  & 61.12\std{6.46} & 60.22\std{4.51} & \textbf{65.93\std{4.36}} & 71.71\std{4.97} & \textbf{71.42\std{3.22}} & 69.75\std{5.16} \\

    \bottomrule
    \end{tabular}
    } 
    \vspace{-4mm}
\end{table*}

%% file: tables/results/ablation_study/non_erp_pretrain_study.tex
\begin{table*}[t]
    \centering
    \caption{\textbf{Effect of Non-ERP EEG Dataset Pretraining (Linear Probing Results).} The \textbf{Non-ERP} denotes pretraining on 5 non-ERP EEG datasets, followed by linear probing on each downstream dataset. The \textbf{Non-ERP \& ERP} jointly pretrains on 38 ERP and 5 non-ERP EEG datasets, whereas \textbf{Non-ERP → ERP} performs sequential non-ERP and ERP pretraining. The \textbf{Supervised} and \textbf{ERP} models are the same as the fully supervised and linear probing results in Table~\ref{tab:erp_pretrain_study}. All results in the table are F1 scores.
    }
    \vspace{-2mm}
    \label{tab:non_erp_pretrain_study_probe}
    \resizebox{1.0\textwidth}{!}{%
    \begin{tabular}{@{}l|c|c|c|c|c|c@{}}
    \toprule
    \multicolumn{1}{l|}{\diagbox{\textbf{Models}}{\textbf{Datasets}}}  & \textbf{CESCA-AODD} & \textbf{CESCA-VODD} &  \textbf{CESCA-FLANKER} &  \textbf{TDBrain-ODD}  & \textbf{NSERP-MSIT} & \textbf{NSERP-ODD} \\
    \midrule
    \multicolumn{1}{l|}{\textbf{Supervised}} & \textbf{53.90\std{0.86}} & \textbf{67.39\std{1.98}} & 64.03\std{1.15} & 86.67\std{0.83} & \textbf{37.79\std{2.25}} & 64.88\std{2.27} \\
    \multicolumn{1}{l|}{\textbf{Non-ERP}} & 51.36\std{0.80} & 65.11\std{1.56} & 63.81\std{1.32} & 86.98\std{1.31} & 36.33\std{2.53} & 64.79\std{3.41} \\
    \multicolumn{1}{l|}{\textbf{Non-ERP \& ERP}}  & 51.74\std{0.68} & 65.00\std{1.59} & \textbf{64.24\std{0.96}} & \textbf{87.00\std{1.17}} & 36.60\std{2.05} & 64.76\std{3.07} \\
    \multicolumn{1}{l|}{\textbf{Non-ERP → ERP}}  & 51.87\std{0.80} & 65.33\std{1.24} & 64.19\std{1.06} & 86.42\std{0.99} & 36.83\std{2.38} & 63.50\std{3.15} \\
    \multicolumn{1}{l|}{\textbf{ERP}}  & 52.30\std{0.67} & 65.00\std{2.12} & 64.03\std{0.55} & 86.79\std{1.26} & 36.26\std{2.01} & \textbf{65.06\std{2.13}} \\
    
    \midrule

    \midrule
    \multicolumn{1}{l|}{\diagbox{\textbf{Models}}{\textbf{Datasets}}}  & \textbf{PD-SIM}  & \textbf{PD-ODD}  & \textbf{ADHD-WMRI}  & \textbf{SCPD}  & \textbf{RLPD}  & \textbf{AOPD}  \\
    \midrule
    \multicolumn{1}{l|}{\textbf{Supervised}} & \textbf{65.20\std{3.74}} & \textbf{68.15\std{3.06}} & 60.97\std{3.88} & 63.25\std{5.25} & 60.07\std{5.40} & 57.29\std{7.37} \\
    \multicolumn{1}{l|}{\textbf{Non-ERP}} & 58.41\std{6.45} & 63.67\std{4.58} & \textbf{62.53\std{5.91}} & 70.06\std{6.77} & 73.11\std{6.62} & 71.09\std{3.03} \\
    \multicolumn{1}{l|}{\textbf{Non-ERP \& ERP}}  & 56.12\std{7.53} & 62.24\std{7.42} & 62.20\std{3.90} & \textbf{74.67\std{2.83}} & \textbf{74.26\std{6.89}} & \textbf{72.43\std{2.47}} \\
    \multicolumn{1}{l|}{\textbf{Non-ERP → ERP}}  & 55.45\std{4.79} & 60.94\std{6.22} & 61.94\std{4.24} & 72.59\std{4.21} & 73.04\std{7.09} & 69.67\std{2.83} \\
    \multicolumn{1}{l|}{\textbf{ERP}}  & 57.26\std{7.83} & 62.10\std{6.43} & 61.60\std{3.11} & 72.83\std{3.23} & 71.06\std{8.11} & 71.89\std{3.51} \\

    \bottomrule
    \end{tabular}
    } 
    \vspace{-3mm}
\end{table*}

\begin{table*}[t]
    \centering
    \caption{\textbf{Effect of Non-ERP EEG Dataset Pretraining (Fine-tuning Results).} All settings are the same as Table~\ref{tab:non_erp_pretrain_study_probe} but change the downstream evaluation step from linear probing to fine-tuning. The \textbf{ERP} model is the \myname base model results reported in Tables~\ref{tab:erp_event_results},  ~\ref{tab:erp_disease_results}, and ~\ref{tab:erp_pretrain_study}. All results are F1 scores.
    }
    \vspace{-2mm}
    \label{tab:non_erp_pretrain_study_finetune}
    \resizebox{1.0\textwidth}{!}{%
    \begin{tabular}{@{}l|c|c|c|c|c|c@{}}
    \toprule
    \multicolumn{1}{l|}{\diagbox{\textbf{Models}}{\textbf{Datasets}}}  & \textbf{CESCA-AODD} & \textbf{CESCA-VODD} &  \textbf{CESCA-FLANKER} &  \textbf{TDBrain-ODD}  & \textbf{NSERP-MSIT} & \textbf{NSERP-ODD} \\
    \midrule
    \multicolumn{1}{l|}{\textbf{Supervised}} & 53.90\std{0.86} & 67.39\std{1.98} & 64.03\std{1.15} & 86.67\std{0.83} & 37.79\std{2.25} & 64.88\std{2.27} \\
    \multicolumn{1}{l|}{\textbf{Non-ERP}} & \textbf{55.14\std{0.92}} & \textbf{70.21\std{1.92}} & 65.38\std{1.21} & 89.82\std{0.62} & 38.48\std{2.85} & \textbf{70.10\std{3.32}} \\
    \multicolumn{1}{l|}{\textbf{Non-ERP \& ERP}}  & 54.43\std{0.44} & 68.66\std{0.76} & \textbf{65.45\std{1.51}} & 90.10\std{1.00} & \textbf{38.78\std{2.67}} & 69.54\std{3.02} \\
    \multicolumn{1}{l|}{\textbf{Non-ERP → ERP}}  & 54.71\std{0.90} & 69.56\std{1.32} & 65.19\std{1.11} & 89.57\std{0.76} & 38.55\std{2.85} & 69.54\std{2.93} \\
    \multicolumn{1}{l|}{\textbf{ERP}}  & 54.61\std{1.09} & 69.75\std{1.44} & 65.44\std{1.11} & \textbf{90.21\std{0.95}} & 38.49\std{2.52} & 68.93\std{2.49} \\
    
    \midrule

    \midrule
    \multicolumn{1}{l|}{\diagbox{\textbf{Models}}{\textbf{Datasets}}}  & \textbf{PD-SIM}  & \textbf{PD-ODD}  & \textbf{ADHD-WMRI}  & \textbf{SCPD}  & \textbf{RLPD}  & \textbf{AOPD}  \\
    \midrule
    \multicolumn{1}{l|}{\textbf{Supervised}} & \textbf{65.20\std{3.74}} & \textbf{68.15\std{3.06}} & 60.97\std{3.88} & 63.25\std{5.25} & 60.07\std{5.40} & 57.29\std{7.37} \\
    \multicolumn{1}{l|}{\textbf{Non-ERP}} & 59.34\std{8.11} & 59.73\std{4.80} & 60.86\std{4.96} & 69.65\std{6.20} & 70.87\std{4.97} & 68.38\std{3.74} \\
    \multicolumn{1}{l|}{\textbf{Non-ERP \& ERP}}  & 58.02\std{6.88} & 58.59\std{3.99} & 63.30\std{3.62} & \textbf{71.95\std{4.61}} & 71.14\std{4.13} & 68.85\std{3.38} \\
    \multicolumn{1}{l|}{\textbf{Non-ERP → ERP}}  & 57.57\std{6.72} & 57.47\std{5.92} & 60.46\std{3.80} & 69.88\std{6.98} & 70.24\std{4.72} & 66.83\std{2.08} \\
    \multicolumn{1}{l|}{\textbf{ERP}}  & 61.12\std{6.46} & 60.22\std{4.51} & \textbf{65.93\std{4.36}} & 71.71\std{4.97} & \textbf{71.42\std{3.22}} & \textbf{69.75\std{5.16}} \\

    \bottomrule
    \end{tabular}
    } 
    \vspace{-3mm}
\end{table*}

%% file: tables/results/ablation_study/patch_length_study.tex
\begin{table*}[t]
    \centering
    \caption{\textbf{Patch Length Study.} The \textbf{50/0.25s} model denotes the patch length $L = 50$ in the \myname embedding, which is 0.25 seconds, as all datasets are resampled to 200Hz. All results are F1 scores.
    }
    \vspace{-2mm}
    \label{tab:patch_length_study}
    \resizebox{1.0\textwidth}{!}{%
    \begin{tabular}{@{}l|c|c|c|c|c|c@{}}
    \toprule
    \multicolumn{1}{l|}{\diagbox{\textbf{Models}}{\textbf{Datasets}}}  & \textbf{CESCA-AODD} & \textbf{CESCA-VODD} &  \textbf{CESCA-FLANKER} &  \textbf{TDBrain-ODD}  & \textbf{NSERP-MSIT} & \textbf{NSERP-ODD} \\
    \midrule
    \multicolumn{1}{l|}{\textbf{200/1.0s}}  & 54.09\std{0.31} & 66.77\std{1.48} & 64.16\std{0.85} & 87.20\std{0.89} & 38.01\std{2.07} & 67.15\std{2.28} \\
    \multicolumn{1}{l|}{\textbf{100/0.5s}}  & 54.48\std{0.46} & 68.57\std{1.64} & 64.95\std{1.10} & 88.49\std{0.90} & \textbf{39.10\std{2.72}} & 67.70\std{2.35} \\
    \multicolumn{1}{l|}{\textbf{50/0.25s (ERP-FM)}}  & \textbf{54.61\std{1.09}} & \textbf{69.75\std{1.44}} & \textbf{65.44\std{1.11}} & \textbf{90.21\std{0.95}} & 38.49\std{2.52} & \textbf{68.93\std{2.49}} \\

    \midrule

    \midrule
    \multicolumn{1}{l|}{\diagbox{\textbf{Models}}{\textbf{Datasets}}}  & \textbf{PD-SIM}  & \textbf{PD-ODD}  & \textbf{ADHD-WMRI}  & \textbf{SCPD}  & \textbf{RLPD}  & \textbf{AOPD}  \\
    \midrule
    \multicolumn{1}{l|}{\textbf{200/1.0s}}  & 58.10\std{6.89} & 58.87\std{5.15} & 62.31\std{5.01} & 69.02\std{5.10} & 68.77\std{5.40} & 67.78\std{5.14} \\
    \multicolumn{1}{l|}{\textbf{100/0.5s}}  & \textbf{61.31\std{7.34}} & \textbf{61.41\std{4.58}} & 64.92\std{3.17} & 69.88\std{6.95} & \textbf{71.79\std{3.63}} & 68.63\std{7.34} \\
    \multicolumn{1}{l|}{\textbf{50/0.25s (ERP-FM)}}  & 61.12\std{6.46} & 60.22\std{4.51} & \textbf{65.93\std{4.36}} & \textbf{71.71\std{4.97}} & 71.42\std{3.22} & \textbf{69.75\std{5.16}} \\
    
    \bottomrule
    \end{tabular}
    } 
    \vspace{-3mm}
\end{table*}

%% file: tables/results/ablation_study/mask_strategy_study.tex
\begin{table*}[t]
    \centering
    \caption{\textbf{Masking Strategy Study.} The \textbf{Random} denotes the use of random masking in pre-training, while \textbf{Mixed} uses random, temporal, and spatial masking strategies together. All results are F1 scores.
    }
    \vspace{-2mm}
    \label{tab:mask_ratio_study}
    \resizebox{1.0\textwidth}{!}{%
    \begin{tabular}{@{}l|c|c|c|c|c|c@{}}
    \toprule
    \multicolumn{1}{l|}{\diagbox{\textbf{Models}}{\textbf{Datasets}}}  & \textbf{CESCA-AODD} & \textbf{CESCA-VODD} &  \textbf{CESCA-FLANKER} &  \textbf{TDBrain-ODD}  & \textbf{NSERP-MSIT} & \textbf{NSERP-ODD} \\
    \midrule
    \multicolumn{1}{l|}{\textbf{Random}}  & 54.53\std{0.80} & 68.94\std{0.85} & 65.11\std{1.05} & 89.65\std{0.90} & \textbf{38.58\std{2.75}} & \textbf{69.01\std{3.21}} \\
    \multicolumn{1}{l|}{\textbf{Mixed (ERP-FM)}}  & \textbf{54.61\std{1.09}} & \textbf{69.75\std{1.44}} & \textbf{65.44\std{1.11}} & \textbf{90.21\std{0.95}} & 38.49\std{2.52} & 68.93\std{2.49} \\
    
    \midrule

    \midrule
    \multicolumn{1}{l|}{\diagbox{\textbf{Models}}{\textbf{Datasets}}}  & \textbf{PD-SIM}  & \textbf{PD-ODD}  & \textbf{ADHD-WMRI}  & \textbf{SCPD}  & \textbf{RLPD}  & \textbf{AOPD}  \\
    \midrule
    \multicolumn{1}{l|}{\textbf{Random}}  & 55.75\std{5.27} & 54.77\std{3.28} & 58.39\std{7.63} & \textbf{76.71\std{3.65}} & \textbf{71.86\std{5.90}} & \textbf{70.14\std{5.40}} \\
    \multicolumn{1}{l|}{\textbf{Mixed (ERP-FM)}}  & \textbf{61.12\std{6.46}} & \textbf{60.22\std{4.51}} & \textbf{65.93\std{4.36}} & 71.71\std{4.97} & 71.42\std{3.22} & 69.75\std{5.16} \\
    
    \bottomrule
    \end{tabular}
    } 
    \vspace{-3mm}
\end{table*}

%% file: tables/results/case_study/trial_averaging_study.tex

\begin{table*}[h]
    \centering
    \caption{\textbf{Trial Averaging Study.} 
    Following conventional ERP analysis, trials with the same subject ID $s$ and cognitive event/condition ID $c$ are \textbf{averaged} before downstream adaptation. We also report \textbf{single-trial} downstream results for comparison. The number of trials under the dataset name denotes the single and averaged trials, respectively. Some datasets either do not provide publicly available event IDs or contain inconsistent event-ID mappings; we omit averaged-trial results for these datasets. The \textbf{(Single, Fine-tuning)} model denotes the \myname base model results reported in Tables~\ref{tab:erp_event_results} and~\ref{tab:erp_disease_results}. All results are F1 scores.
    }
    \vspace{-2mm}
    \label{tab:trial_averaging_study}
    \resizebox{1.0\textwidth}{!}{%
    \begin{tabular}{@{}ll|c|c|c|c|c|c@{}}
    \toprule
    \textbf{Trial Setting} & \diagbox{\textbf{Models}}{\textbf{Datasets}}
    & \makecell{\textbf{CESCA-AODD} \\ \textit{(38,151/228)} }
    & \makecell{\textbf{CESCA-VODD}  \\ \textit{(20,419/196)} }
    & \makecell{\textbf{CESCA-FLANKER}  \\ \textit{(29,774/146)} }
    & \makecell{\textbf{TDBrain-ODD}  \\ \textit{(54,400/254)} }
    & \makecell{\textbf{NSERP-MSIT} \\ \textit{(16,729/168)} }
    & \makecell{\textbf{NSERP-ODD} \\ \textit{(27,865/126)} }
    \\
    \midrule
    \multirow{3}{*}{\textbf{Single}} & \textbf{Supervised} & 53.90\std{0.86} & 67.39\std{1.98} & 64.03\std{1.15} & 86.67\std{0.83} & 37.79\std{2.25} & 64.88\std{2.27} \\
    & \textbf{Linear Probing} & 52.30\std{0.67} & 65.00\std{2.12} & 64.03\std{0.55} & 86.79\std{1.26} & 36.26\std{2.01} & 65.06\std{2.13} \\
    & \textbf{Fine-tuning} & 54.61\std{1.09} & 69.75\std{1.44} & 65.44\std{1.11} & 90.21\std{0.95} & 38.49\std{2.52} & 68.93\std{2.49} \\
    \midrule
    \multirow{3}{*}{\textbf{Averaged}} & \textbf{Supervised} & 75.77\std{18.76} & 82.62\std{5.03} & \textbf{81.80\std{7.86}} & \textbf{100.00\std{0.00}} & 45.04\std{9.53} & 80.39\std{2.82} \\
    & \textbf{Linear Probing} & \textbf{97.39\std{0.87}} & 93.97\std{4.95} & 75.20\std{10.14} & 99.62\std{0.77} & 50.20\std{12.78} & 79.62\std{7.62} \\
    & \textbf{Fine-tuning} & 96.94\std{2.97} & \textbf{96.99\std{2.46}} & 75.62\std{8.97} & \textbf{100.00\std{0.00}} & \textbf{54.44\std{6.62}} & \textbf{86.51\std{5.27}} \\

    \midrule

    \midrule
    \textbf{Trial Setting} & \diagbox{\textbf{Models}}{\textbf{Datasets}}
    & \makecell{\textbf{PD-SIM}  \\ \textit{(55,921/\textbf{---})} }
    & \makecell{\textbf{PD-ODD}  \\ \textit{(34,702/\textbf{---})} }
    & \makecell{\textbf{ADHD-WMRI}  \\ \textit{(21,832/\textbf{---})} }
    & \makecell{\textbf{SCPD}  \\ \textit{(46,193/550)} }
    & \makecell{\textbf{RLPD}  \\ \textit{(21,510/334)} }
    & \makecell{\textbf{AOPD}  \\ \textit{(14,625/150)} }
    \\
    \midrule
    \multirow{3}{*}{\textbf{Single}} & \textbf{Supervised} & \textbf{65.20\std{3.74}} & \textbf{68.15\std{3.06}} & 60.97\std{3.88} & 63.25\std{5.25} & 60.07\std{5.40} & 57.29\std{7.37} \\
    & \textbf{Linear Probing} & 57.26\std{7.83} & 62.10\std{6.43} & 61.60\std{3.11} & 72.83\std{3.23} & 71.06\std{8.11} & 71.89\std{3.51} \\
    & \textbf{Fine-tuning} & 61.12\std{6.46} & 60.22\std{4.51} & \textbf{65.93\std{4.36}} & 71.71\std{4.97} & 71.42\std{3.22} & 69.75\std{5.16} \\
    \midrule
    \multirow{3}{*}{\textbf{Averaged}} & \textbf{Supervised} & \textbf{---} & \textbf{---} & \textbf{---} & 63.91\std{6.63} & 59.08\std{4.80} & 58.78\std{13.14} \\
    & \textbf{Linear Probing} & \textbf{---} & \textbf{---} & \textbf{---} & \textbf{88.78\std{5.52}} & 77.42\std{4.02} & \textbf{76.61\std{8.87}} \\
    & \textbf{Fine-tuning} & \textbf{---} & \textbf{---} & \textbf{---} & 84.43\std{5.31} & \textbf{79.14\std{3.58}} & 75.48\std{7.20} \\

    \bottomrule
    \end{tabular}
    } 
    \vspace{-3mm}
\end{table*}

%% file: tables/results/case_study/disease_detection_study.tex
\begin{table*}[h]
    \centering
    \caption{\textbf{Subject-Level Neurological Disease Detection.} This table compares different training strategies for neurological disease detection, reporting both trial-level performance and subject-level performance after majority voting. The number of trials under the dataset name denotes the single and averaged trials, respectively. The terms \textbf{Single} and \textbf{Averaged} follow the same definitions as in Table~\ref{tab:trial_averaging_study}. 
    }
    \vspace{-2mm}
    \label{tab:disease_detection_study}
    \resizebox{\textwidth}{!}{
    \begin{tabular}{@{}ll|ccc|ccc|ccc@{}}
    \toprule

    \multicolumn{2}{c|}{\textbf{Metrics}} & \textbf{Accuracy} & \textbf{F1 Score} & \textbf{AUROC} & \textbf{Accuracy} & \textbf{F1 Score} & \textbf{AUROC} & \textbf{Accuracy} & \textbf{F1 Score} & \textbf{AUROC} \\ 
    \midrule

    \textbf{Trial Setting} & \diagbox{\textbf{Models}}{\textbf{Datasets}}
    & \multicolumn{3}{c|}{\makecell{\textbf{SCPD} \\ \textit{(46,193/550 Trials, 55 Subjects)} } }
    & \multicolumn{3}{c|}{\makecell{\textbf{RLPD} \\ \textit{(21,510/334 Trials, 56 Subjects)} } }
    & \multicolumn{3}{c}{\makecell{\textbf{AOPD}  \\ \textit{(14,625/150 Trials, 50 Subjects)} } }
    \\ 
    \midrule

    \multicolumn{2}{c}{} & \multicolumn{9}{c}{\textbf{Trial-Level Classification}}  \\
    \midrule

    \multirow{3}{*}{\textbf{Single}} 
    & \textbf{Supervised}  & 66.82\std{5.54} & 63.25\std{5.25} & 72.65\std{7.35} & 64.34\std{4.99} & 60.07\std{5.40} & 67.41\std{7.80} & 59.89\std{9.03} & 57.29\std{7.37} & 61.63\std{9.02}  \\
    & \textbf{Linear Probing}  & 77.62\std{1.40} & 72.83\std{3.23} & 82.84\std{4.48} & 73.99\std{8.12} & 71.06\std{8.11} & 81.49\std{11.22} & 75.81\std{2.85} & 71.89\std{3.51} & 79.28\std{5.96}  \\
    & \textbf{Fine-tuning}  & 76.65\std{4.76} & 71.71\std{4.97} & 81.04\std{4.71} & 74.81\std{2.96} & 71.42\std{3.22} & 79.55\std{5.76} & 74.32\std{4.91} & 69.75\std{5.16} & 79.27\std{8.35}  \\
    \midrule
    \multirow{3}{*}{\textbf{Averaged}} 
    & \textbf{Supervised}  & 65.00\std{5.94} & 63.91\std{6.63} & 71.19\std{8.74} & 59.56\std{4.87} & 59.08\std{4.80} & 64.28\std{6.29} & 60.67\std{11.23} & 58.78\std{13.14} & 63.91\std{8.54}  \\
    & \textbf{Linear Probing}  & \textbf{88.83\std{5.47}} & \textbf{88.78\std{5.52}} & \textbf{95.49\std{3.61}} & 77.52\std{4.03} & 77.42\std{4.02} & 85.52\std{3.80} & \textbf{77.33\std{8.27}} & \textbf{76.61\std{8.87}} & \textbf{81.42\std{6.74}}  \\
    & \textbf{Fine-tuning}  & 84.67\std{5.21} & 84.43\std{5.31} & 95.24\std{4.12} & \textbf{79.20\std{3.57}} & \textbf{79.14\std{3.58}} & \textbf{89.13\std{3.97}} & 76.67\std{5.96} & 75.48\std{7.20} & 78.58\std{4.19}  \\

    \midrule
    \multicolumn{2}{c}{} & \multicolumn{9}{c}{\textbf{Subject-Level Detection}}  \\
    \midrule

    \multirow{3}{*}{\textbf{Single}} 
    & \textbf{Supervised}  & 60.00\std{9.72} & 58.79\std{9.52} & 77.22\std{12.96} & 63.33\std{11.30} & 61.50\std{12.54} & 76.67\std{13.56} & 62.00\std{16.00} & 60.02\std{15.40} & 66.40\std{16.89}  \\
    & \textbf{Linear Probing}  & 78.33\std{4.08} & 77.54\std{4.57} & 85.28\std{4.44} & 75.00\std{11.79} & 73.20\std{14.46} & 85.00\std{12.12} & 74.00\std{10.20} & 73.36\std{10.58} & 81.20\std{7.11}  \\
    & \textbf{Fine-tuning}  & 76.67\std{6.24} & 75.71\std{6.64} & 83.89\std{3.24} & 73.33\std{6.24} & 72.77\std{6.37} & 83.33\std{5.83} & 70.00\std{8.94} & 68.91\std{9.20} & 86.80\std{6.65}  \\
    \midrule
    \multirow{3}{*}{\textbf{Averaged}} 
    & \textbf{Supervised}  & 68.33\std{6.24} & 66.54\std{6.96} & 71.11\std{14.15} & 63.33\std{10.00} & 62.35\std{9.93} & 68.61\std{10.82} & 66.00\std{16.25} & 62.21\std{20.46} & 66.80\std{20.42}  \\
    & \textbf{Linear Probing}  & \textbf{91.67\std{5.27}} & \textbf{91.54\std{5.42}} & \textbf{98.61\std{2.15}} & 90.00\std{6.24} & 89.88\std{6.33} & 95.28\std{4.86} & 90.00\std{10.95} & 89.37\std{12.05} & 94.40\std{8.52}  \\
    & \textbf{Fine-tuning}  & 90.00\std{8.16} & 89.71\std{8.40} & 96.11\std{4.76} & \textbf{95.00\std{4.08}} & \textbf{94.97\std{4.11}} & \textbf{98.89\std{1.62}} & \textbf{92.00\std{11.66}} & \textbf{91.39\std{12.79}} & \textbf{96.00\std{8.00}}  \\

    \bottomrule
    \end{tabular}
    }
\vspace{-2mm}
\end{table*}

%% file: 070conclusion.tex
\textbf{Conclusion.} In this work, we introduced \myname, to the best of our knowledge, the first foundation model specifically developed for ERP representation learning. Built on a large-scale corpus of 1,517,157 single-trial ERPs from 3,696 subjects across 38 datasets and 18 paradigms, \myname achieves the best overall average rank among 18 evaluated methods on 12 downstream datasets spanning ERP event/condition classification and neurological disease classification. Our analyses further show that both ERP and non-ERP EEG pretraining can provide transferable representations for ERP tasks, while fine-grained temporal tokenization is important for effectively modeling transient ERP dynamics. In addition, single-trial and averaged-trial ERP exhibit complementary advantages: single-trial ERP provides the data scale and variability needed for large-scale self-supervised pretraining, whereas averaged-trial ERP provides cleaner downstream signals with improved SNR. Combining single-trial pretraining with averaged-trial downstream adaptation therefore provides an effective strategy, particularly for subject-level neurological disease detection.

\textbf{Limitations.} Despite these encouraging results, several limitations remain. First, the benefit of pretraining is not consistent across all downstream datasets, indicating that the effectiveness of pretrained ERP representations can still depend on dataset characteristics and downstream tasks. More broadly, no single method, including \myname, consistently achieves the best performance across all datasets and evaluation metrics; although our model obtains the best overall average rank, substantial room remains for developing more robust and universally transferable ERP representations. Second, performance under single-trial analysis remains relatively limited for both fully supervised learning and pretrained downstream adaptation, reflecting the difficulty of decoding low-SNR ERP responses without trial averaging. Improving single-trial representation learning and downstream decoding therefore remains an important direction for future work. Finally, this study focuses primarily on predictive performance and does not provide dedicated interpretability analyses. Future work could investigate how learned representations relate to established ERP characteristics, including waveform morphology, latency, spatial patterns, and task- or disease-related neurophysiological differences. Overall, our findings establish a promising foundation-model framework for ERP analysis while highlighting several open challenges toward more robust, interpretable, and generalizable ERP representation learning.

%% file: 080appendix.tex
\input{tables/datasets/erp_pretraining_datasets}

\input{tables/datasets/non_erp_pretraining_datasets}

\section{Pretraining Datasets}
\label{sec:pretraining_datasets}

We list all the ERP and non-ERP pretraining datasets in Table~\ref{tab:erp_pretraining_data} and ~\ref{tab:non_erp_pretraining_data}, respectively. For convenience, in addition to the ERP/EEG task, preprocessing parameters, and statistical information, we also provide the reference paper and the raw data download link in this table for future ERP research.

\section{Reproducibility and Implementation Details}
\label{sec:implementation_details}

To ensure reproducibility, we implement and evaluate all methods under the unified training and evaluation framework described in Section~\ref{sub:experiment_setup}. The source code, model configurations, and complete running scripts for \myname and all baseline methods are provided in our project repository. All datasets used in this study are publicly accessible through the sources reported in the dataset tables. Unless otherwise specified below, the shared optimization, early stopping, cross-validation, and evaluation settings follow Section~\ref{sub:experiment_setup}. For the 4 EEG foundation-model baselines, we initialize the models from their officially released pretrained checkpoints and fine-tune the pretrained backbone together with the newly added downstream components. We provide the model-specific implementation details below.

\textbf{EEG Features.}
The EEG Features baseline follows the generic handcrafted EEG features used in ERP-Benchmark~\citep{wang2026benchmarking}. For each EEG channel, we extract 31 features covering four groups: 10 time-domain statistical features (mean, standard deviation, variance, skewness, kurtosis, interquartile range, median, minimum, maximum, and root-mean-square amplitude); 11 power-related features (absolute delta, theta, alpha, and beta power, total power, theta-to-alpha and alpha-to-beta ratios, and relative delta, theta, alpha, and beta power); 7 spectral-shape features (spectral centroid, spectral roll-off, peak frequency, peak power, mean frequency, median frequency, and spectral flatness); and 3 entropy-related features (normalized spectral entropy, Tsallis entropy, and spectral entropy). Spectral quantities are computed using Welch power spectral density estimation. Features from all channels are flattened and fed into a linear classifier.

\textbf{ERP Features.}
The ERP Features baseline uses ERP-oriented handcrafted features following ERP-Benchmark~\citep{wang2026benchmarking}. For each channel, we extract 91 features describing ERP waveform morphology, spectral characteristics, and temporal dynamics. Specifically, temporal pyramid pooling at levels $\{1,2,4,8\}$ divides each ERP epoch into multiple temporal segments, from which mean, standard deviation, root-mean-square amplitude, line length, and peak-to-peak amplitude are extracted, resulting in 75 features. We additionally compute positive and negative peak amplitudes and their normalized latencies, and 8 spectral features, including relative band powers, total power, spectral centroid, spectral flatness, and median frequency, normalized spectral entropy, and the three Hjorth parameters. The resulting features from all channels are flattened and passed to a linear classifier.

\textbf{TCN}~\citep{bai2018empirical} is a temporal convolutional architecture based on stacked dilated convolutional blocks for modeling long-range temporal dependencies. We use 6 hidden convolutional stages with 128 hidden channels, followed by a 320-dimensional output representation, with kernel size $=3$. Global max pooling over the temporal dimension is applied before the final linear classifier.

\textbf{ModernTCN}~\citep{luo2024moderntcn} is a convolutional architecture designed for general time-series representation learning using large-kernel depthwise convolutions and pointwise channel mixing. We use \textit{patch\_len} $=32$, \textit{stride} $=16$, \textit{num\_blocks} $=[1,1,1]$, \textit{large\_size} $=[9,9,9]$, \textit{small\_size} $=[5,5,5]$, \textit{dims} $=[32,64,128]$, and \textit{ffn\_ratio} $=1$.

\textbf{TimesNet}~\citep{wu2023timesnet} transforms one-dimensional time series into two-dimensional representations according to dominant temporal periodicities and applies 2D convolution for feature extraction. We use 2 TimesBlocks with \textit{top\_k} $=1$, \textit{d\_model} $=32$, \textit{d\_ff} $=64$, and 6 convolutional kernels in each Inception block.

\textbf{PatchTST}~\citep{nie2022time} adopts channel-independent temporal patching and Transformer encoding for time-series representation learning. We use 6 Transformer encoder layers with \textit{n\_heads} $=8$, \textit{d\_model} $=128$, and \textit{d\_ff} $=256$. The temporal patch length is set to 100 samples with a stride of 8 samples. The encoded representations from all channels and temporal patches are flattened and passed to a linear classifier.

\textbf{iTransformer}~\citep{liuitransformer} reverses the conventional temporal-token formulation by embedding each time-series variable as an independent token and modeling dependencies across variables through self-attention. We use 6 Transformer encoder layers with \textit{n\_heads} $=8$, \textit{d\_model} $=128$, and \textit{d\_ff} $=256$. The output tokens from all EEG channels are concatenated for final classification.

\textbf{Medformer}~\citep{wang2024medformer} is designed for biomedical time-series classification and employs multi-granularity patch embedding together with intra- and inter-granularity attention. We use 6 encoder layers with \textit{n\_heads} $=8$, \textit{d\_model} $=128$, and \textit{d\_ff} $=256$. Three temporal granularities with \textit{patch\_len\_list} $=[25,50,100]$ are used, with the stride of each granularity set equal to its patch length. Inter-granularity attention is enabled.

\textbf{MedGNN}~\citep{fan2025towards} is a graph-based model for medical time-series classification that jointly models temporal representations at multiple resolutions and dynamic inter-variable relationships. We use four temporal resolutions with \textit{resolution\_list} $=[2,4,6,8]$ and \textit{nodedim} $=10$. Both the difference-attention and multi-resolution Transformer modules use 6 encoder layers with \textit{n\_heads} $=8$, \textit{d\_model} $=128$, and \textit{d\_ff} $=256$, followed by the multi-resolution graph neural network and a linear classification head.

\textbf{EEGNet}~\citep{lawhern2018eegnet} is a compact convolutional architecture designed for EEG decoding using temporal and spatial convolutions. Our implementation first applies temporal convolution and depthwise temporal refinement with 16 feature maps, followed by depthwise spatial convolution across EEG channels and pointwise projection to 32 feature maps. A second depthwise-separable convolution further mixes local spatial information. Adaptive pooling produces a fixed 128-dimensional representation, which is fed into a linear classifier.

\textbf{EEGInception}~\citep{zhang2021eeg} uses multi-scale temporal convolutions to extract EEG patterns at different temporal resolutions together with spatial feature extraction. We stack 3 Inception-spatial blocks with output channel dimensions $(96,192,384)$. The temporal convolution kernel sizes are $(8,16,32)$, the bottleneck dimension is 32, and the spatial depth multiplier is 2. Global average pooling is applied before the final classifier.

\textbf{EEGConformer}~\citep{song2022eeg} combines convolutional EEG feature extraction with Transformer self-attention to jointly capture local and global dependencies. We first use its shallow convolutional embedding module and then apply 6 Transformer encoder layers with \textit{n\_heads} $=8$, \textit{d\_model} $=128$, and \textit{d\_ff} $=256$. The resulting token representations are flattened and passed to a linear classifier.

\textbf{EEGDeformer}~\citep{ding2024eeg} combines convolutional EEG embedding with hierarchical coarse-to-fine Transformer modeling and dense information extraction. At the unified 200~Hz sampling rate, we use a temporal convolution kernel of 21 samples, corresponding to approximately 0.1~s. The convolutional embedding uses 128 feature kernels, followed by up to 6 hierarchical coarse-to-fine Transformer layers with \textit{n\_heads} $=8$ and feed-forward dimension $=256$. Features from the hierarchical Transformer and the dense information branches are concatenated for final classification.

\textbf{BIOT}~\citep{yang2024biot} is a foundation model for heterogeneous biosignal representation learning based on single-channel spectral tokenization. We initialize the model from the officially released EEG-PREST checkpoint pretrained for 16-channel EEG. Because the downstream ERP datasets contain heterogeneous channel configurations, we prepend a learnable $1\times1$ Conv1D channel-mapping layer that projects the original channels to 16 channels. The pretrained BIOT backbone uses an embedding dimension of 256, 8 attention heads, and 4 Transformer layers, with an FFT window of 200 samples and hop length of 100 samples. The channel-mapping layer, pretrained backbone, and downstream classifier are jointly optimized during fine-tuning. We use the released checkpoint \texttt{EEG-PREST-16-channels.ckpt}.

\textbf{LaBraM}~\citep{jiang2024large} is a large EEG foundation model pretrained using neural tokenization and masked EEG modeling. We initialize LaBraM from its officially released base checkpoint. To accommodate heterogeneous downstream channel configurations, a learnable $1\times1$ Conv1D layer maps each dataset to the 128-channel input representation expected by the released model implementation. Each signal is divided into 200-sample temporal patches with a stride of 100 samples, corresponding to 1-second patches with 50\% overlap at 200~Hz. The backbone uses a 200-dimensional embedding, 12 Transformer layers, 10 attention heads, and an MLP ratio of 4. The original classification head is replaced with a new task-specific linear classifier, and the entire model is fine-tuned on each downstream dataset. We use the officially released \texttt{labram-base.pth} checkpoint.

\textbf{CBraMod}~\citep{wang2024cbramod} is an EEG foundation model based on a criss-cross Transformer that separately models spatial and temporal dependencies. We initialize the model from its officially released pretrained checkpoint. A learnable $1\times1$ Conv1D layer maps the original EEG channels of each dataset to the 19-channel configuration used during CBraMod pretraining. Input sequences are zero-padded to a multiple of 200 samples and divided into 200-sample temporal segments. The pretrained backbone uses 12 layers, 8 attention heads, \textit{d\_model} $=200$, and a feed-forward dimension of 800. We remove the original output projection and attach a task-specific classification head consisting of two hidden layers with dimensions 800 and 200 before the final classifier. The channel adapter, pretrained backbone, and classifier are jointly fine-tuned on each downstream dataset. We use the officially released \texttt{pretrained\_weights.pth} checkpoint.

\textbf{REVE}~\citep{el2026reve} is a large-scale EEG foundation model designed to adapt to heterogeneous electrode configurations through explicit electrode-position information. We initialize the model from the released REVE-Base checkpoint (\texttt{brain-bzh/reve-base}). Unlike BIOT, LaBraM, and CBraMod, we do not map the input to a fixed channel configuration. Instead, we retain the original channels of each downstream dataset and obtain their 3D electrode positions from the corresponding MNE standard montage using the channel metadata provided by our dataset loader. The EEG signals and electrode coordinates are jointly passed to the pretrained REVE backbone. We mean-pool its token-level representations when necessary and use a task-specific classification head consisting of LayerNorm, dropout, and a linear projection. The pretrained backbone and classifier are jointly fine-tuned on each downstream dataset.

\textbf{\myname.}
Our base \myname uses a 12-layer Transformer encoder and a lightweight 2-layer Transformer decoder, with \textit{d\_model} $=128$, \textit{n\_heads} $=8$, \textit{d\_ff} $=256$, and dropout $=0.1$, for a total of around 2.4 million parameters. Both the encoder and decoder use full self-attention, RMSNorm, and SwiGLU feed-forward blocks. We use non-overlapping single-channel patches with \textit{patch\_len} $=50$, corresponding to 0.25~s at the unified sampling rate of 200~Hz. During self-supervised pretraining, the masking ratio is set to 0.5, and one of random, temporal, and spatial masking is randomly selected for each mini-batch. Smooth L1 loss with $\beta=0.5$ is applied only to the masked patches. During downstream adaptation, the decoder and reconstruction head are discarded, and a linear classifier is applied to the flattened output of all encoder tokens. For linear probing, the patch embedding and encoder are frozen, and only the classifier is optimized, whereas fine-tuning jointly updates the pretrained encoder and classifier. We release the pretrained checkpoint in our code repository.

%% file: tables/datasets/erp_pretraining_datasets.tex
\begin{table*}[t]
    \centering
    \caption{\textbf{ERP Pretraining Dataset Statistics.} All the ERP pretraining datasets follow the same preprocessing pipeline, including a band-pass filter, artifact removal, and resampling to 200Hz. The epoch and baseline window depend on the ERP tasks and datasets, following the reference paper of each dataset if provided. For three HBN-EEG datasets, we use publicly available releases 1 to 11. Abbreviations: \textbf{AODD}: Auditory Oddball; \textbf{VODD}: Visual Oddball; \textbf{MSIT+}: Extended multi-source interference task; \textbf{SIM}: Simon Conflict; \textbf{RL}: Reinforcement Learning; \textbf{SJT}: Semantic Judgment Task; \textbf{PST}: Probability Selection Task; \textbf{VS}: Visual Search; \textbf{VD}: Visual Discrimination; \textbf{CCD}: Contrast Change Detection; \textbf{SUS}: Surround Suppression; \textbf{SYS}: Symbol Search; \textbf{DPX}: Dot Probe Expectancy Task; \textbf{VWM}: Visual Working Memory Task; \textbf{SCT}: Semantic Categorization Task; \textbf{SIM-RL}: Simon Conflict with Reinforcement Learning.
    }
    \vspace{-2mm}
    \label{tab:erp_pretraining_data}
    \resizebox{1.0\textwidth}{!}{%
    \begin{tabular}{@{}l|ccccccc@{}}
    \toprule
    \multicolumn{1}{l|}{\textbf{Datasets}}  & \textbf{ERP Task} & \textbf{\#Subjects} &  \textbf{Baseline(s)} &  \textbf{Epoch(s)}  & \textbf{\#Trials} & \textbf{\#Channels} & \textbf{Raw Data Link}  \\ 
    \midrule
    \multicolumn{1}{l|}{\textbf{AUD-MAB}~\citep{campbell2023electrophysiological}} & RL & 49 &  [-0.5, 0.0]  & [-0.5, 1.0] & 14,828 & 19 & \href{https://openneuro.org/datasets/ds005907/versions/1.0.0}{Openneuro} \\
    \multicolumn{1}{l|}{\textbf{AUD-PS}~\citep{singh2023affective}} & RL & 50 &  [-0.5, 0.0]  & [-0.3, -0.2] & 8,000 & 59 & \href{https://openneuro.org/datasets/ds004515/versions/1.0.0}{Openneuro} \\
    \multicolumn{1}{l|}{\textbf{AVSPP}~\citep{chailloux2020single}} & VODD & 50 &  [-0.2, 0.0]  & [-0.2, 1.0] & 75,048 & 8 & \href{https://openneuro.org/datasets/ds003190/versions/1.0.1}{Openneuro} \\
    \multicolumn{1}{l|}{\textbf{AVSS-AODD}~\citep{vceponiene2008modality}} & AODD & 49 &  [-0.2, 0.0]  & [-0.2, 0.8] & 68,885 & 33 & \href{https://openneuro.org/datasets/ds002893/versions/2.0.0}{Openneuro} \\
    \multicolumn{1}{l|}{\textbf{AVSS-VODD}~\citep{vceponiene2008modality}} & VODD & 49 &  [-0.2, 0.0]  & [-0.2, 0.8] & 68,805 & 33 & \href{https://openneuro.org/datasets/ds002893/versions/2.0.0}{Openneuro} \\
    \multicolumn{1}{l|}{\textbf{CCT-MJAH}~\citep{schwartz2023neurophysiological}} & SJT & 31 &  [-0.2, 0.0]  & [-0.2, 0.8] & 3,327 & 32 & \href{https://openneuro.org/datasets/ds004860/versions/1.0.0}{Openneuro} \\
    \multicolumn{1}{l|}{\textbf{EPSD}~\citep{cavanagh2019multiple}} & PST & 122 &  [-0.2, 0.0]  & [-0.2, 0.8] & 42,690 & 60 & \href{https://openneuro.org/datasets/ds003474/versions/1.1.0}{Openneuro} \\
    \multicolumn{1}{l|}{\textbf{ERPCORE-ERN}~\citep{kappenman2021erp}} & Flanker & 38 &  [-0.4, -0.2]  & [-0.6, 0.4] & 16,048 & 30 & \href{https://osf.io/q6gwp/files/osfstorage}{OSF} \\
    \multicolumn{1}{l|}{\textbf{ERPCORE-LRP}~\citep{kappenman2021erp}} & Flanker & 38 &  [-0.8, -0.6]  & [-0.8, 0.2] & 16,048 & 30 & \href{https://osf.io/28e6c/files/osfstorage}{OSF} \\
    \multicolumn{1}{l|}{\textbf{ERPCORE-MMN}~\citep{kappenman2021erp}} & AODD & 40 &  [-0.2, 0.0]  & [-0.2, 0.8] & 39,232 & 30 & \href{https://osf.io/5q4xs/files/osfstorage}{OSF} \\
    \multicolumn{1}{l|}{\textbf{ERPCORE-N2pc}~\citep{kappenman2021erp}} & VS & 38 &  [-0.2, 0.0]  & [-0.2, 0.8] & 12,807 & 30 & \href{https://osf.io/yefrq/files/osfstorage}{OSF} \\
    \multicolumn{1}{l|}{\textbf{ERPCORE-N170}~\citep{kappenman2021erp}} & VD & 40 &  [-0.2, 0.0]  & [-0.2, 0.8] & 12,800 & 30 & \href{https://osf.io/pfde9/files/osfstorage}{OSF} \\
    \multicolumn{1}{l|}{\textbf{ERPCORE-N400}~\citep{kappenman2021erp}} & SJT & 38 &  [-0.2, 0.0]  & [-0.2, 0.8] & 4,800 & 30 & \href{https://osf.io/29xpq/files/osfstorage}{OSF} \\
    \multicolumn{1}{l|}{\textbf{ERPCORE-P3}~\citep{kappenman2021erp}} & VODD & 39 &  [-0.2, 0.0]  & [-0.2, 0.8] & 8,000 & 30 & \href{https://osf.io/etdkz/files/osfstorage}{OSF} \\
    \multicolumn{1}{l|}{\textbf{Go-Nogo}~\citep{fabre2001limit}} & Go-Nogo & 19 &  [-0.1, 0.0]  & [-0.5, 1.0] & 34,992 & 19 & \href{https://openneuro.org/datasets/ds002680/versions/1.0.0}{Openneuro} \\
    \multicolumn{1}{l|}{\textbf{HBN-EEG-CCD}~\citep{shirazi2024hbn}} & CCD & 1,966 &  [-0.2, 0.0]  & [-0.2, 0.8] & 136,107 & 129 & \href{https://openneuro.org/datasets/ds005516/versions/1.0.1}{Openneuro} \\
    \multicolumn{1}{l|}{\textbf{HBN-EEG-SUS}~\citep{shirazi2024hbn}} & SUS & 2,297 &  [-0.2, 0.0]  & [-0.2, 0.8] & 264,958 & 129 & \href{https://openneuro.org/datasets/ds005516/versions/1.0.1}{Openneuro} \\
    \multicolumn{1}{l|}{\textbf{HBN-EEG-SYS}~\citep{shirazi2024hbn}} & SYS & 1,562 &  [-0.2, 0.0]  & [-0.2, 0.8] & 26,601 & 129 & \href{https://openneuro.org/datasets/ds005516/versions/1.0.1}{Openneuro} \\
    \multicolumn{1}{l|}{\textbf{HeartBEAM}~\citep{kim2025electroencephalography}} & VODD & 68 &  [-0.2, 0.0]  & [-0.5, 1.0] & 64,645 & 65 & \href{https://openneuro.org/datasets/ds006480/versions/1.0.0}{Openneuro} \\
    \multicolumn{1}{l|}{\textbf{IMS-ODD}~\citep{goldman2020improvisation}} & AODD & 28 &  [-0.1, 0.0]  & [-0.4, 1.6] & 36,149 & 64 & \href{https://openneuro.org/datasets/ds003570/versions/1.0.0}{Openneuro} \\
    \multicolumn{1}{l|}{\textbf{MMPST-EXP1-TRAIN}~\citep{jackson2023reduced}} & PST & 50 &  [-0.2, 0.0]  & [-0.2, 0.8] & 11,949 & 59 & \href{https://openneuro.org/datasets/ds004315/versions/1.0.0}{Openneuro} \\
    \multicolumn{1}{l|}{\textbf{MMPST-EXP1-TEST}~\citep{jackson2023reduced}} & PST & 50 &  [-0.2, 0.0]  & [-0.2, 0.8] & 6,000 & 59 & \href{https://openneuro.org/datasets/ds004315/versions/1.0.0}{Openneuro} \\
    \multicolumn{1}{l|}{\textbf{MMPST-EXP2-TRAIN}~\citep{jackson2023reduced}} & PST & 50 &  [-0.2, 0.0]  & [-0.2, 0.8] & 11,939 & 59 & \href{https://openneuro.org/datasets/ds004315/versions/1.0.0}{Openneuro} \\
    \multicolumn{1}{l|}{\textbf{MMPST-EXP2-TEST}~\citep{jackson2023reduced}} & PST & 49 &  [-0.2, 0.0]  & [-0.2, 0.8] & 5,884 & 59 & \href{https://openneuro.org/datasets/ds004315/versions/1.0.0}{Openneuro} \\
    \multicolumn{1}{l|}{\textbf{MRI-AODD}~\citep{delorme2022eeg}} & AODD & 13 &  [-0.2, 0.0]  & [-0.2, 0.8] & 9,706 & 64 & \href{https://openneuro.org/datasets/ds003061/versions/1.1.2}{Openneuro} \\
    \multicolumn{1}{l|}{\textbf{mTBI-DPX}~\citep{cavanagh2020joint}} & DPX & 91 &  [-0.3, -0.2]  & [-0.5, 1.0] & 53,937 & 61 & \href{https://openneuro.org/datasets/ds005114/versions/1.0.0}{Openneuro} \\
    \multicolumn{1}{l|}{\textbf{mTBI-ODD}~\citep{cavanagh2019erps}} & AODD & 96 &  [-0.2, 0.0]  & [-0.2, 0.8] & 50,362 & 61 & \href{https://openneuro.org/datasets/ds003522/versions/1.1.0}{Openneuro} \\
    \multicolumn{1}{l|}{\textbf{mTBI-VWM}~\citep{broadway2019executive}} & VWM & 91 &  [-0.2, 0.0]  & [-0.2, 0.8] & 32,588 & 61 & \href{https://openneuro.org/datasets/ds003523/versions/1.1.0}{Openneuro} \\
    \multicolumn{1}{l|}{\textbf{NAFPS}~\citep{lee2025neural}} & SCT & 24 &  [-0.1, 0.0]  & [-0.1, 1.6] & 7,199 & 29 & \href{https://openneuro.org/datasets/ds005565/versions/1.0.3}{Openneuro} \\
    \multicolumn{1}{l|}{\textbf{PLAF-EXP1}~\citep{brown2022reward}} & RL & 25 &  [-0.2, 0.0]  & [-0.2, 0.8] & 5,000 & 63 & \href{https://openneuro.org/datasets/ds003822/versions/1.1.0}{Openneuro} \\
    \multicolumn{1}{l|}{\textbf{PLAF-EXP2}~\citep{brown2022reward}} & RL & 25 &  [-0.2, 0.0]  & [-0.2, 0.8] & 3,997 & 59 & \href{https://openneuro.org/datasets/ds003753/versions/1.1.0}{Openneuro} \\
    \multicolumn{1}{l|}{\textbf{PSTCC}~\citep{cavanagh2014conflict}} & PST & 110 &  [-0.2, 0.0]  & [-0.2, 0.8] & 44,313 & 63 & \href{https://openneuro.org/datasets/ds004532/versions/1.2.0}{Openneuro} \\
    \multicolumn{1}{l|}{\textbf{Runabout}~\citep{liebherr2021eeg}} & AODD & 39 &  [-0.2, 0.0]  & [-0.2, 0.8] & 176,941 & 32 & \href{https://openneuro.org/datasets/ds003620/versions/1.1.1}{Openneuro} \\
    \multicolumn{1}{l|}{\textbf{SICE}~\citep{toffolo2022evoking}} & SJT & 24 &  [-0.2, 0.0]  & [-0.2, 0.8] & 9,600 & 128 & \href{https://datadryad.org/dataset/doi:10.5061/dryad.6wwpzgmx4}{Dryad} \\
    \multicolumn{1}{l|}{\textbf{SIMCC-TRAIN}~\citep{cavanagh2014conflict}} & SIM-RL & 110 &  [-0.3, -0.2]  & [-0.5, 1.0] & 74,400 & 63 & \href{https://openneuro.org/datasets/ds003518/versions/1.1.0}{Openneuro} \\
    \multicolumn{1}{l|}{\textbf{SIMCC-TEST}~\citep{cavanagh2014conflict}} & SIM-RL & 110 &  [-0.3, -0.2]  & [-0.5, 1.0] & 39,455 & 63 & \href{https://openneuro.org/datasets/ds003518/versions/1.1.0}{Openneuro} \\
    \multicolumn{1}{l|}{\textbf{TABG}~\citep{cavanagh2015cortical}} & RL & 23 &  [-0.3, -0.2]  & [-0.5, 1.0] & 11,040 & 63 & \href{https://openneuro.org/datasets/ds003458/versions/1.1.0}{Openneuro} \\
    \multicolumn{1}{l|}{\textbf{VWMCC}~\citep{broadway2018dopamine}} & VWM & 27 &  [-0.2, 0.0]  & [-0.2, 0.8] & 8,077 & 63 & \href{https://openneuro.org/datasets/ds003519/versions/1.1.0}{Openneuro} \\

    \bottomrule
    \end{tabular}
    } 
    \vspace{-2mm}
\end{table*}

%% file: tables/datasets/non_erp_pretraining_datasets.tex
\begin{table*}[t]
    \centering
    \caption{\textbf{Non-ERP EEG Pretraining Datasets Statistics.} All the Non-ERP EEG pretraining datasets follow the same preprocessing pipeline, including a band-pass filter, artifact removal, and resampling to 200Hz. The only difference compared with the ERP dataset is the absence of baseline correction. Abbreviations: \textbf{RS}: Resting State; \textbf{PS}: Photic Stimulation; \textbf{HV}: Hyperventilation; 
    }
    \vspace{-2mm}
    \label{tab:non_erp_pretraining_data}
    \resizebox{1.0\textwidth}{!}{%
    \begin{tabular}{@{}l|ccccccc@{}}
    \toprule
    \multicolumn{1}{l|}{\textbf{Datasets}}  & \textbf{EEG Task} & \textbf{\#Subjects} &  \textbf{Baseline(s)} &  \textbf{Epoch(s)}  & \textbf{\#Trials} & \textbf{\#Channels} & \textbf{Raw Data Link}  \\ 
    \midrule
    \multicolumn{1}{l|}{\textbf{BACA-RS}~\citep{gajewski2022impact}} & RS & 608  & --- & [0.0, 1.0] & 611,269 & 64 & \href{https://openneuro.org/datasets/ds005385/versions/1.0.3}{Openneuro} \\
    \multicolumn{1}{l|}{\textbf{CAUEEG}~\citep{kim2023deep}} & RS \& PS \& HV & 1,379  & --- & [0.0, 2.0] & 507,892 & 64 & \href{https://github.com/ipis-mjkim/caueeg-dataset}{Github} \\
    \multicolumn{1}{l|}{\textbf{P-ADIC}~\citep{shor2021eeg}} & RS & 249  & --- & [0.0, 2.0] & 135,621 & 19 & \href{https://datadryad.org/dataset/doi:10.5061/dryad.8gtht76pw}{Dryad} \\
    \multicolumn{1}{l|}{\textbf{TDBrain}~\citep{van2022two}} & RS & 1,273  & --- & [0.0, 2.0] & 160,937 & 26 & \href{https://brainclinics.com/resources}{Brainclinics} \\
    \multicolumn{1}{l|}{\textbf{TUEP}~\citep{veloso2017big}} & RS & 200  & --- & [0.0, 2.0] & 1,129,171 & 19 & \href{https://isip.piconepress.com/projects/nedc/html/tuh_eeg/index.shtml}{TUH EEG} \\

    \bottomrule
    \end{tabular}
    } 
    \vspace{-5mm}
\end{table*}